\documentclass[12pt,draftclsnofoot,onecolumn]{IEEEtran}

\IEEEoverridecommandlockouts
\usepackage{indentfirst,flushend}
\usepackage{amssymb,bm,mathrsfs,times,amscd,amsmath,amsthm,amsfonts,bbm}
\usepackage{latexsym}
\usepackage{dsfont}
\usepackage{graphicx}
\usepackage{subfigure}
\usepackage{color}
\usepackage{cases}
\newtheorem{theorem}{Theorem}
\newtheorem{corollary}{Corollary}

\newtheorem{lemma}{Lemma}
\newtheorem{remark}{Remark}
\usepackage{multirow}
\usepackage[sort]{cite}
\usepackage{multicol}
\usepackage{clrscode}
\usepackage{algorithm}
\usepackage{psfrag,stfloats,nicefrac}
\usepackage{algorithmic}
\usepackage{supertabular}
\usepackage{makecell}
\usepackage{booktabs}
\usepackage{threeparttable}
\usepackage[justification=centering]{caption}
\allowdisplaybreaks[4]

\begin{document}

\title{Beamforming Design for Multiuser uRLLC with Finite Blocklength Transmission}
\author{Shiwen~He,~Zhenyu An,~Jianyue Zhu,~Jian Zhang,\\~Yongming Huang,~and Yaoxue~Zhang
\thanks{S. He and J. Zhang are with the School of Computer Science and Engineering, Central South University, Changsha 410083, China. S. He is also with the Purple Mountain Laboratories, Nanjing 210096, China. (email: \{shiwen.he.hn,jianzhang\}@csu.edu.cn). }
\thanks{Z. An is with the Purple Mountain Laboratories, Nanjing 210096, China. (email: anzhenyu\_155@126.com). }
\thanks{J. Zhu is with the School of Information Science and Engineering, Southeast University, Nanjing 210096, China.  She is also with the College of Electronic and Information Engineering, Nanjing University of Information Science and Technology, Nanjing, China. (email: zhujy@seu.edu.cn). }
\thanks{Y. Huang is with the National Mobile Communications Research Laboratory, School of Information Science and Engineering, Southeast University, Nanjing 210096, China. He is also with the Purple Mountain Laboratories, Nanjing 210096, China. (email: huangym@seu.edu.cn). }
\thanks{Y. Zhang is with the Department of Computer Science and Technology, Tsinghua University, Beijing 100084, China. (email: zhangyx@tsinghua.edu.cn)}
}

\maketitle
\vspace{-.6 in}

\begin{abstract}
Driven by the explosive growth of Internet of Things (IoT) devices with stringent requirements on latency and reliability, ultra-reliability and low latency communication (uRLLC) has become one of the three key communication scenarios for the fifth generation (5G) and beyond 5G communication systems. In this paper, we focus on the beamforming design problem for the downlink multiuser uRLLC systems. Since the strict demand on the reliability and latency, in general, short packet transmission is a favorable form for uRLLC systems, which indicates the literature Shannon's capacity formula is no longer applicable. With the finite blocklength transmission, the achievable delivery rate is greatly influenced by the reliability and latency. Using the developed achievable delivery rate formula for finite blocklength transmission, we respectively formulate the problems of interest as the weighted sum rate maximization, energy efficiency maximization, and user fairness optimization by considering the maximum allowable transmission power and minimum rate requirement. It's worthy pointing out that this is the first work to design the beamforming vectors for the downlink multiuser uRLLC systems. To address these non-convex problems, some important insights have been discovered by analyzing the function of achievable delivery rate. For example, the minimum rate requirement can be realized by low bounded the signal-to-interference-plus-noise ratio. Based on the discovered results, we provide algorithms to optimize the beamforming vectors and power allocation, which are guaranteed to converge to a local optimum solution of the formulated problems with low computational complexity. Our simulation results reveal that our proposed beamforming algorithms outperform the zero-forcing beamforming algorithm with equal power allocation widely used in the existing literatures.
\end{abstract}

\section{ Introduction}

Driven by the explosive growth of Internet of Thing (IoT) devices with stringent requirements on latency and reliability, ultra-reliability and low latency communication (uRLLC) has become a part of the emerging 5th generation (5G) and beyond 5G communication systems. uRLLC is one of the most innovative technical schemes for the 5G mobile network. Practically, the reliability requirement of the factory automation and tele-surgery is $1-10^{-9}$ with end-to-end latency being less than $1$ ms~\cite{8631201}. Other services, e.g., smart grids, intelligent transportation systems, and process automation, have more relaxed reliability requirements of $\left(1-10^{-3}\right)\sim(1-10^{-6})$ at latencies between $1$ ms to $100$ ms. The 3rd generation partnership project (3GPP) proposed the general uRLLC requirement of $99.999\%$ (block error rate (BLER) of $10^{-5}$) for one transmission of a packet, which is with $32$ bytes and $1$ ms latency~\cite{3GPP}.

The existing study on coordinated multiuser communication is carried based on the classical Shannon capacity~\cite{BellShannon1948,JSACHe2019}, which assumes having an arbitrarily low decoding error probability for very long transmission blocklengths. However, in practical uRLLC systems, a large number of IoT devices have to communicate using short bursts of data. In other words, finite blocklength (short-packet) transmission is the typical form for uRLLC systems, e.g., measurements and control commands for industrial manufacturing and control systems are $10$ to $30$ bytes~\cite{7733543}. As a result, the literature Shannon's capacity formula is no longer applicable in uRLLC systems due to the finite blocklength transmission~\cite{8933345}. For uRLLC systems, the decoding error probability is non-negligible due to the impact of the finite blocklength transmission. In a word, three key performance indicators of reliability, latency, and throughput have a fundamental correlation for uRLLC systems. Specifically, in uRLLC systems, the performance improvement can be realized from three dimensions, i.e., directly reducing latency, directly increasing reliability, and improving the throughput via reducing latency and increasing reliability~\cite{8636206}. The authors in~\cite{shirvanimoghaddam2019short} provided an overview of channel coding techniques for uRLLC systems and compared them in terms of performance and complexity.

In recent years, the short-packet communication was widely investigated. The achievable rate of finite blocklength transmission is investigated in~\cite{TITPoly2010}, which is smaller than the Shannon rate, with taking into account the decoding error probability, the number of transmitting bits, and the finite blocklength (channels used). The authors in~\cite{TITErseghe2016} discussed the application of the asymptotic uniform expansion approach for evaluating the achievable bounds in the finite blocklength regime. The authors in~\cite{TWCSun2018} introduced the downlink non-orthogonal multiple access (NOMA) into short-packet communications, where the effective throughput of the user with a higher channel gain was maximized. A wireless-powered IoTs network with short-packet communication was studied in~\cite{TWCChen2019}. The authors in~\cite{TWCZheng2020} analyzed the impact of finite blocklength secrecy coding on the design of secure transmissions in slow fading channel. The instantaneous and average block error was investigated for the simultaneous information and power transmission  uRLLC relay system~\cite{TCOMAaga2020}.

Note that the achievable rate for uRLLC systems is a complicated function and hence the resource allocation problem for uRLLC systems is tricky. In the literature, significant attention has been devoted to studying and developing resource allocation algorithms enabling uRLLC, e.g.,~\cite{8402240,8253477,8638800,arXivGhanem2019}. In~\cite{8402240}, in a downlink multiuser network operating with finite blocklength codes, the authors proposed optimal power allocation algorithms to maximize the normalized sum throughput under QoS constraints. The uplink-downlink bandwidth configuration and delay components were jointly optimized to minimize the total bandwidth~\cite{8253477}. In~\cite{8638800}, the authors studied the resource allocation problem for ultra-reliable low-latency edge computing, where the users\textquoteright{} power consumption was minimized. In~\cite{arXivGhanem2019}, the authors studied the resource allocation for downlink orthogonal frequency division multiple access (OFDMA) systems with the objective of maximizing weighted system sum rate for uRLLC systems. In~\cite{8705373}, the authors discussed the principles of wireless access for uRLLC systems, where massive MIMO and multi-connectivity were considered. In~\cite{9187217}, the authors studied the problem of dynamic channel allocation for uRLLC systems while assuming the absence of instantaneous channel state information (CSI) at the transmitter. The authors in~\cite{TWCXu2016} formulated an energy-efficient packet scheduling problem by adopting the channel capacity formula for finite blocklength.


Though a large number of works on the uRLLC systems have been carried on in the last few years, to the best of our knowledge, there are no existing works investigating the beamforming design problem for uRLLC multi-antenna multiuser systems. Nevertheless, the application of multi-antenna in uRLLC systems is necessary for further improving the quality of experiences of user and enhancing the delivery data rate. In this paper, we focus on the beamforming design for multiuser uRLLC systems and the contributions are listed as follows.
\begin{itemize}
    \item We first analysis the property of the achievable rate for uRLLC systems to obtain some key insights. For example, the function of the achievable rate is a non-convex and non-concave function with respect to the signal-to-interference-pus-noise ratio (SINR).
    \item The feasibility of the considered problems is studied. Practically, if the rate of one of the users approaches zero, the delivery latency is infinite for the user, which is undesirable for uRLLC systems and leads to the infeasibility of the formulated problems. To avoid the infinite delivery latency, we obtain an analytical condition under which the rate of any user can satisfy its minimum rate requirement and hence the feasibility of the problem is guaranteed.
    \item We investigate the beamforming design respectively for weighted sum rate maximization, energy efficiency maximization, and user fairness optimization in uRLLC systems by considering the maximum allowable transmission power and minimum rate requirement.
    \item Algorithmic solutions to the formulated problems are provided by using some basic mathematical operations, successive convex approximation method, and the uplink-downlink duality theory. 
    \item We further discuss the initialization of the provided algorithms by solving the formulated power minimization problem, which greatly improves the efficiency of beamforming design.
     \item Finally, the proposed solution is evaluated via simulation. It's found that our proposed beamforming design outperforms the classical zero-forcing beamforming (ZFBF) algorithm with equal power allocation.
\end{itemize}

The rest of the paper is organized as follows. Section \ref{introduction} describes
the uRLLC system model and the formulated problem. Section \ref{function property} analyzes the properties of the achievable rate of uRLLC systems. Section \ref{beamforming design} studies
the  beamforming design respectively for sum rate maximization and energy efficiency maximization. The performance of the proposed beamforming solution is shown in Section \ref{simulation} via simulation. Finally, Section \ref{conclusion} concludes this paper.

\textbf{\textcolor{black}{$\mathbf{\mathit{Notations}}$}}: We respectively use lower case letters and boldface capital to denote vectors and matrices. $\boldsymbol{a}^{H}$ denotes the Hermitian transpose vector $\boldsymbol{a}$; $\left|.\right|$ and $\left\Vert.\right\Vert$ respectively denote the absolute value of a complex scalar and the Euclidean vector norm; $\mathbb{C}^{T}$ denotes the set of complex numbers;

\section{\label{introduction}System Model and Problem Formulation}

In this section, we present the system model of the downlink multiuser uRLLC system, including a multi-antenna base station (BS) and $K$ single antenna users. The BS is assumed to be equipped with $N_{\mathrm{t}}$ transmitting antennas and $K\leq N_{\mathrm{t}}$. For simplicity, let $\mathcal{K}=\left\{1,2,\cdots,K\right\}$ be the set of users.
\begin{figure}[t]
\renewcommand{\captionfont}{\footnotesize}
	\centering
	\includegraphics[width=0.6\columnwidth,keepaspectratio]{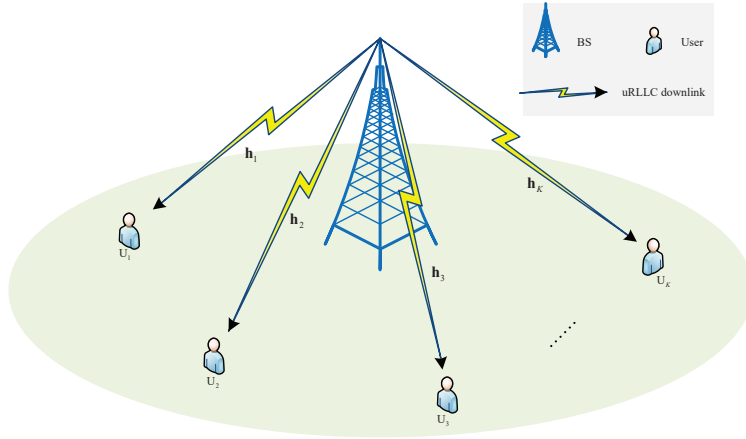}\\
	\caption{System model of the downlink multiuser uRLLC system.}
	\label{EquivalentSystemModel}
\end{figure}
Consequently, the received baseband signal at the $k$-th user is given by:
\begin{equation}\label{URLLC01}
y_{k}=\sum\limits_{l\in\mathcal{K}}\sqrt{p_{l}}\mathbf{h}_{k}^{H}\mathbf{w}_{l}s_{l}+n_{k},
\end{equation}
where $\mathbf{w}_{k}\in\mathbb{C}^{N_{\mathrm{t}}\times 1}$ and $p_{k}$ denote the normalized beamforming vector and transmitting power used by the BS for the $k$-th user, respectively. $\mathbf{h}_{k}\in\mathbb{C}^{N_{\mathrm{t}}\times 1}$ represents the channel coefficient between the BS and the $k$-th user. $s_{k}$ represents the baseband signal for the $k$-th user and $n_{k}$ denotes the additive white Gaussian noise with $\mathcal{CN}\left(0,\sigma_{k}^{2}\right)$ at the $k$-th user. Thus, the signal-to-interference-plus-noise ratio (SINR) of the $k$-th user is calculated as~\eqref{URLLC02} with $\overline{\mathbf{h}}_{k}=\frac{\mathbf{h}_{k}}{\sigma_{k}}$.
\begin{equation}\label{URLLC02}
\gamma_{k}=\frac{p_{k}\left|\overline{\mathbf{h}}_{k}^{H}\mathbf{w}_{k}\right|^{2}}
{\sum\limits_{l\neq k}p_{l}\left|\overline{\mathbf{h}}_{k}^{H}\mathbf{w}_{l}\right|^{2}+1}.
\end{equation}

Generally speaking, Shannon's formula adopted by the existing literature is based on the assumption of the blocklength approaching infinity and the decoding error probability going to zero~\cite{BellShannon1948}. Namely, it has no ability to capture the relation among the achievable rate, decoding error probability, and transmission latency for the communication system with a finite blocklength $n$~\cite{TITPoly2010}. Here, the blocklength denotes the codeword length or the number of channels used. Therefore, in the finite blocklength regime which
is restricted to a finite number of channels used, we need to look for a new method to characterize this relation. Fortunately, as can be seen in~\cite[Fig.~1]{TITErseghe2016} and~\cite[E4.277]{PhDPolyanskiy}, the relation among the achievable rate, decoding error probability, transmission latency, and finite blocklength $n$ can be described as
\begin{equation}\label{URLLC03}
R\left(\gamma_{k}\right)=C\left(\gamma_{k}\right)-\vartheta\sqrt{V\left(\gamma_{k}\right)},
\end{equation}
where $C\left(\gamma\right)=\ln\left(1+\gamma\right)$,  $\vartheta=\frac{Q^{-1}\left(\epsilon\right)}{\sqrt{n}}$ with $\epsilon$ being a desirable decoding error probability, $Q^{-1}\left(\cdot\right)$ being the inverse of Gaussian Q-function $Q\left(x\right)=\frac{1}{\sqrt{2\pi}}\int_{x}^{\infty}\mathrm{exp}\left(-\frac{t^{2}}{2}\right)dt$,  
and $V\left(\gamma\right)$ is defined as
\begin{equation}\label{URLLC04}
V\left(\gamma\right)=1-\frac{1}{\left(1+\gamma\right)^{2}}.
\end{equation}
Different from Shannon capacity, the achievable rate defined in~\eqref{URLLC03} takes the error probability into account, i.e., the second item in the right-hands of~\eqref{URLLC03}.

Generally specking, the spectral efficiency, energy efficiency, and user fairness are three key indices of performance metrics for wireless communication systems. Meanwhile, a general uRLLC reliability requirement for the transmission of a packet is $10^{-5}$ for $32$ bytes (i.e. the number of transmitting data bits  is $D=256$ bits) with a user plane latency of $1$ ms~\cite{8636206}. Motivated by these observations, in this paper, we focus on investigating these three problems for multiuser uRLLC communication systems by considering the maximum allowable transmission power and minimum rate requirement. Firstly, the weighted sum rate maximization (SRMax) problem is formulated as
\begin{subequations}\label{URLLC05}
\begin{align}
&\max_{\left\{\mathbf{w}_{k}, p_{k}\right\}}~\sum\limits_{k\in\mathcal{K}}\alpha_{k}R\left(\gamma_{k}\right),\label{URLLC05a}\\
\mathrm{s.t.}~&\sum\limits_{k\in\mathcal{K}}p_{k}\leq P, \left\|\mathbf{w}_{k}\right\|_{2}^{2}=1, \forall k\in\mathcal{K}, \label{URLLC05b}\\
&\frac{D}{n}\ln\left(2\right)\leq R\left(\gamma_{k}\right), \forall k\in\mathcal{K}. \label{URLLC05c}
\end{align}
\end{subequations}
where $\alpha_{k}$ denotes the prior of the $k$-th user. Secondly, the problem of energy efficiency maximization (EEMax) is formulated as
\begin{subequations}\label{URLLC06}
\begin{align}
&\max_{\left\{\mathbf{w}_{k}, p_{k}\right\}}~
\frac{\sum\limits_{k\in\mathcal{K}}\alpha_{k}R\left(\gamma_{k}\right)}
{\eta\sum\limits_{k\in\mathcal{K}}p_{k}+N_{\mathrm{t}}P_{\mathrm{c}}+P_{0}},\label{URLLC06a}\\
\mathrm{s.t.}~&\sum\limits_{k\in\mathcal{K}}p_{k}\leq P, \left\|\mathbf{w}_{k}\right\|_{2}^{2}=1, \forall k\in\mathcal{K}, \label{URLLC06b}\\
&\frac{D}{n}\ln\left(2\right)\leq R\left(\gamma_{k}\right), \forall k\in\mathcal{K}. \label{URLLC06c}
\end{align}
\end{subequations}
where $\eta\geq 1$ is a constant which accounts for the inefficiency of the power amplifier. $P_{\mathrm{c}}$ is the constant circuit power consumption per-antenna including power dissipations in the transmit filter, mixer, frequency synthesizer, and digital-to-analog converter, which are independent of the actual transmitting power. $P_{0}$ is the basic power consumed at the BS and is independent of the number of transmitting antennas~\cite{TSPHe2014}. Another focus of the uRLLC system is to ensure that the worst user rate is maximized. Namely, the fairness optimization (MaxMin) problem is formulated as follows
\begin{subequations}\label{URLLC07}
\begin{align}
&\max_{\left\{\mathbf{w}_{k}, p_{k}\right\}}\min\limits_{k\in\mathcal{K}}~R\left(\gamma_{k}\right),\label{URLLC07a}\\
\mathrm{s.t.}~&\sum\limits_{k\in\mathcal{K}}p_{k}\leq P, \left\|\mathbf{w}_{k}\right\|_{2}^{2}=1, \forall k\in\mathcal{K}, \label{URLLC07b}\\
&\frac{D}{n}\ln\left(2\right)\leq R\left(\gamma_{k}\right), \forall k\in\mathcal{K}. \label{URLLC07c}
\end{align}
\end{subequations}
Note that in this paper, by properly designing the beamforming vectors, the number of bits received by the $k$-th user with decoding error probability $\epsilon_{k}$ can be determined. It is well known that the weighted SRMax, EEMax, and fairness (maxmin) problems, i.e.,~\eqref{URLLC05},~\eqref{URLLC06}, and~\eqref{URLLC07}, are non-convex and hard to obtain their solutions for interference channel communication systems. Furthermore, compared to Shannon formula, the rate expression defined in~\eqref{URLLC03} is more complex, therefore, problems~\eqref{URLLC05},~\eqref{URLLC06}, and~\eqref{URLLC07} are also non-convex and difficult to directly obtain their solutions.

\section{\label{function property} Property Analysis of Function $R\left(\gamma\right)$}

In this section, we focus on analyzing function $R\left(\gamma\right)=C\left(\gamma\right)-\vartheta\sqrt{V\left(\gamma\right)}$ in the range $\gamma\geq 0$ to obtain some key insights. Note that if the achievable rate of one of the users approaches zero, then the delivery latency is infinite for the user, i.e., the system latency of multiuser communication system is also infinite. However, this is undesirable for the uRLLC system. On the other hand, the user rate is limited by the maximum allowable transmission power $P$. Based on this observation, we assume that $0<\gamma_{k}\leq\tilde{\gamma}_{k}=\frac{P\left\|\mathbf{h}_{k}^{H}\right\|^{2}}{\sigma_{k}^{2}}$, $\forall k\in\mathcal{K}$, where $\tilde{\gamma}_{k}$ denotes the SINR of the $k$-th user which is obtained via maximum ratio transmission with full power transmission and without suffering any interference. Generally speaking, the desirable decoding error probability $\epsilon$ should be at least smaller than $0.5$. In particular, in uRLLC scenarios, the desirable decoding error probability $\epsilon$ should be less than $10^{-6}$~\cite{8636206}. Note that $Q\left(0\right)=0.5$, i.e., $Q^{-1}\left(\epsilon=0.5\right)=0$, and $Q\left(x\right)$ is a decreasing function with respect to $x$. Therefore, without loss of generality, we assume that $Q^{-1}\left(\epsilon\right)>0$.
\begin{theorem}\label{URLLCTheorem01}
Given $\vartheta>0$, function $R\left(\gamma\right)$ is a monotonic decreasing function for $0\leq\gamma<\nu_{0}$ and is a monotonic increasing function for $\gamma>\nu_{0}$,
where $\nu_{0}$ is given by
\begin{equation}\label{URLLC08}
\nu_{0}=\sqrt{\frac{1+\sqrt{1+4\vartheta^{2}}}{2}}-1,
\end{equation}
and $R\left(\nu_{0}\right)\leq0$.
\end{theorem}
\begin{IEEEproof}
The first and second order derivation of function $R\left(\gamma\right)$ are calculated as follows:
\begin{subequations}\label{URLLC09}
\begin{align}
R'\left(\gamma\right)&=\frac{1}{1+\gamma}\left(1-\vartheta\frac{1}{\left(1+\gamma\right)\sqrt{\left(1+\gamma\right)^{2}-1}}\right),\label{URLLC09a}\\
R''\left(\gamma\right)
&=-\frac{1}{\left(1+\gamma\right)^{2}}\left[1-\frac{\vartheta\left(3\left(1+\gamma\right)^{2}-2\right)}
{\left(1+\gamma\right)\left(\left(1+\gamma\right)^{2}-1\right)^{\frac{3}{2}}}\right].\label{URLLC09b}
\end{align}
\end{subequations}
Let $R'\left(\gamma\right)=0$, we have
\begin{equation}\label{URLLC10}
\left(1+\gamma\right)\sqrt{\left(1+\gamma\right)^{2}-1}=\vartheta,
\end{equation}
After some basic mathematic operation,  the solution to~\eqref{URLLC10} is given as:
\begin{equation}\label{URLLC11}
\nu_{0}=\sqrt{\frac{1+\sqrt{1+4\vartheta^{2}}}{2}}-1,
\end{equation}
In addition, note that $R''\left(\nu_{0}\right)>0$. This implies that $\nu_{0}$ is an extreme small value point of function $R\left(\gamma\right)$. Note that function $\left(1+\gamma\right)\sqrt{\left(1+\gamma\right)^{2}-1}$ is a monotonic increasing function with respect to $\gamma>0$. Therefore, we have
\begin{enumerate}
\item When $0\leq\gamma<\nu_{0}$, we have $R'\left(\gamma\right)<0$, i.e., function $R\left(\gamma\right)$ is a monotonic decreasing function.
\item When $\nu_{0}<\gamma$, we have $R'\left(\gamma\right)>0$, i.e., function $R\left(\gamma\right)$ is a monotonic increasing function.
\end{enumerate}
Recalling $R\left(0\right)=0$, we have $R\left(\nu_{0}\right)\leq0$.
\end{IEEEproof}

\begin{remark}
From Theorem \ref{URLLCTheorem01},  it is not difficult to find that $R\left(\gamma\right)=0$ with $\vartheta>0$ has two solutions. One of the two solutions is $\nu_{1}=0$. Note that  $\gamma$ should be larger than the positive solution of $R\left(\gamma\right)=0$, which avoids the rate of one of users is less than or equal to zero. But, it is very hard to find the non-zero analytical solution of $R\left(\gamma\right)$ due to its complex expression. In the sequel, we investigate the solution of $R\left(\gamma\right)=\alpha$ with $\alpha\geq 0$ and $\vartheta>0$ in the range $\gamma>0$ to obtain the non-zero solution $\nu_{2}$ of $R\left(\gamma\right)=0$.
\end{remark}
\begin{theorem}\label{URLLCTheorem02}
When $\vartheta>0$ is fixed, the solution of $R\left(\gamma\right)=\alpha$ with $\alpha\geq 0$ in the range $\gamma>0$ is $\gamma^{*}=e^{\alpha+\frac{\kappa^{*}}{2}}-1$, where $\kappa^{*}=\mathcal{W}\left(^{2\vartheta,-2\vartheta};-4e^{-2\alpha}\vartheta^{2}\right)$ with
$\mathcal{W}\left(^{\iota_{1},\iota_{2}};\mu\right)$ being given by
\begin{equation}\label{URLLC12}
\mathcal{W}\left(^{\iota_{1},\iota_{2}};\mu\right)=\iota_{1}-\sum\limits_{m=1}^{\infty}\frac{1}{m*m!}\left(\frac{\mu m e^{-\iota_{1}}}{\iota_{2}-\iota_{1}}\right)^{m}\mathcal{B}_{m-1}\left(\frac{-2}{\iota_{2}-\iota_{1}}\right), 
\end{equation}
with $\mathcal{B}_{m}\left(z\right)$ being defined as
\begin{equation}\label{URLLC13}
\mathcal{B}_{m-1}\left(z\right)=\sum\limits_{k=0}^{m}\frac{\left(m+k\right)!}{k!\left(m-k\right)!}\left(\frac{z}{2}\right)^{k}.
\end{equation}
\end{theorem}
\begin{IEEEproof}
Rewriting $R\left(\gamma\right)=\alpha$ as follows
\begin{equation}\label{URLLC14}
\ln\left(e^{-\alpha}\left(1+\gamma\right)\right)-\vartheta\sqrt{V\left(\gamma\right)}=0.
\end{equation}
Let $e^{-\alpha}=\beta$ and $\ln\left(\beta\left(1+\gamma\right)\right)=\eta$, then, we have
\begin{equation}\label{URLLC15}
\left(1+\gamma\right)^{-2}=\beta^{2}e^{-2\eta}.
\end{equation}
In addition, according to the range $\gamma>0$ and~\eqref{URLLC14}, it is not difficult to find that $0<\eta\leq\vartheta$. Furthermore, we can rewrite~\eqref{URLLC15} as follows
\begin{equation}\label{URLLC16}
\eta=\vartheta\sqrt{1-\beta^{2}e^{-2\eta}}.
\end{equation}
After some basic mathematical operations,~\eqref{URLLC16} can be rewritten as
\begin{equation}\label{URLLC17}
-\frac{1}{\beta^{2}\vartheta^{2}}\left(\eta-\vartheta\right)\left(\eta+\vartheta\right)=e^{-2\eta}.
\end{equation}
belonging to the class of transcendent equation (generalized Lambert $W$ function)~\cite{ACMMaignan2016,ArXivMun2015,AMSMezo2017}. Let $\kappa=2\eta$, $0<\kappa\leq2\vartheta$, we can reformulate~\eqref{URLLC17} as follows
\begin{equation}\label{URLLC18}
e^{\kappa}\left(\kappa-2\vartheta\right)\left(\kappa+2\vartheta\right)=-4\beta^{2}\vartheta^{2}.
\end{equation}
Note that $-4\beta^{2}\vartheta^{2}<0$, recalling $0<\kappa\leq2\vartheta$, therefore, the solution of~\eqref{URLLC18} belongs to $0<\kappa\leq2\vartheta$. Using the Lagrange inversion theorem, the solution of~\eqref{URLLC18} is given by
\begin{equation}\label{URLLC19}
\kappa^{*}=\mathcal{W}\left(^{2\vartheta,-2\vartheta};-4\beta^{2}\vartheta^{2}\right),
\end{equation}
where $\mathcal{W}\left(^{\iota_{1},\iota_{2}};\mu\right)$ is defined as follows
\begin{equation}\label{URLLC20}
\mathcal{W}\left(^{\iota_{1},\iota_{2}};\mu\right)=\iota_{1}-\sum\limits_{m=1}^{\infty}\frac{1}{m*m!}\left(\frac{\mu m e^{-\iota_{1}}}{\iota_{2}-\iota_{1}}\right)^{m}\mathcal{B}_{m-1}\left(\frac{-2}{m\left(\iota_{2}-\iota_{1}\right)}\right),
\end{equation}
with $\mathcal{B}_{m}\left(z\right)$ being defined as
\begin{equation}\label{URLLC21}
\mathcal{B}_{m}\left(z\right)=\sum\limits_{k=0}^{m}\frac{\left(m+k\right)!}{k!\left(m-k\right)!}\left(\frac{z}{2}\right)^{k}.
\end{equation}
According to~\eqref{URLLC19}, it is not difficult to see that the summation term in equation~\eqref{URLLC20} is a nonnegative number, i.e., $0<\kappa^{*}\leq2\vartheta$. Thus, the solution of $R\left(\gamma\right)=\alpha$ is given by
\begin{equation}\label{URLLC22}
\gamma^{*}= e^{\alpha+\eta^{*}}-1=e^{\alpha+\frac{\kappa^{*}}{2}}-1.
\end{equation}
The proof is completed.
\end{IEEEproof}

\begin{remark}
Note that $R\left(\gamma\right)=0$ and $R\left(\gamma\right)=\frac{D}{n}\ln\left(2\right)$ are two special cases of $R\left(\gamma\right)=\alpha$ with $\alpha=0$ and $\alpha=\frac{D}{n}\ln\left(2\right)$, respectively. This implies that their solutions can be easily obtained by using the conclusion obtained in Theorem~\ref{URLLCTheorem02}. In particular, another solution $\nu_{2}$ of $R\left(\gamma\right)=0$ is given by
\begin{equation}\label{URLLC23}
\nu_{2}= e^{\frac{\kappa^{*}}{2}}-1.
\end{equation}
While, the solution of $R\left(\gamma\right)=\frac{D}{n}\ln\left(2\right)$ is calculated as
\begin{equation}\label{URLLC24}
\nu_{3}= e^{\frac{D}{n}\ln\left(2\right)+\frac{\kappa^{*}}{2}}-1.
\end{equation}
Therefore, to guarantee the feasibility of problems~\eqref{URLLC05}, \eqref{URLLC06}, and \eqref{URLLC07}, the SINR $\gamma_{k}$ has to be larger than $\nu_{3}$, for $\forall k\in\mathcal{K}$, i.e., we have the following Corollary.
\end{remark}
\begin{corollary}\label{corollary1}
To guarantee the feasibility of problems~\eqref{URLLC05},~\eqref{URLLC06}, and~\eqref{URLLC07}, we must have $\tilde{\gamma}_{k}\geq\gamma_{k}\geq\nu_{3}, \forall k\in\mathcal{K}$.
\end{corollary}
\begin{IEEEproof}
This can be easily observed according to the minimum rate constraint and the result provided in Theorem \ref{URLLCTheorem02}.
\end{IEEEproof}

Table~\ref{SolutionEvaluations} illustrates the accuracy of the solution of $R\left(\gamma\right)=\alpha$ obtained by Theorem~\ref{URLLCTheorem02} for different values of $\vartheta$ and $\alpha$, respectively, where $\varpi\left(\gamma\right)=R(\gamma)-\alpha$ and the exact solution (Exact) is obtained by directly using the function \emph{vpasolve}$\left(\cdot\right)$ in Matlab software. Note that when the value of $\vartheta$ is smaller than $0.096$, the approximation error of the solution is slightly large for $R\left(\gamma\right)=0$, especially when the value of $\vartheta$ is close to zero. However, for $\alpha\neq 0$, the approximation error of the solution is very small, even can be ignored.
\begin{table*}[ht]
  \centering
  \resizebox{\textwidth}{!}{
  \begin{threeparttable}
  \caption{\bf Accuracy analysis of solution obtained by Theorem 2 for $R(\gamma)=\alpha$.}\label{SolutionEvaluations}
    \begin{tabular}{cccccccccccccccc}
    \toprule[1.5pt]
    \multicolumn{16}{c}{$R(\gamma)=0$} \cr
    \cmidrule[0.5pt](lr){1-16}
    $\vartheta$ &$\nu_{2}$ &Exact &$\varpi\left(\nu_{2}\right)$ &$\vartheta$ &$\nu_{2}$ &Exact &$\varpi\left(\nu_{2}\right)$ &$\vartheta$ &$\nu_{2}$ &Exact &$\varpi\left(\nu_{2}\right)$ &$\vartheta$ &$\nu_{2}$ &Exact &$\varpi\left(\nu_{2}\right)$ \cr
    \cmidrule[1pt](lr){1-4} \cmidrule[1pt](lr){5-8} \cmidrule[1pt](lr){9-12} \cmidrule[1pt](lr){13-16}

    \textcolor[rgb]{ 1,  0,  0}{\textbf{0.001}} & \textcolor[rgb]{ 1,  0,  0}{\textbf{0.000022 }} & \textcolor[rgb]{ 1,  0,  0}{\textbf{0.000002 }} & \textcolor[rgb]{ 1,  0,  0}{\textbf{1.55E-05}} & \textcolor[rgb]{ 1,  0,  0}{\textbf{0.126}} & \textcolor[rgb]{ 1,  0,  0}{\textbf{0.031294 }} & \textcolor[rgb]{ 1,  0,  0}{\textbf{0.031273 }} & \textcolor[rgb]{ 1,  0,  0}{\textbf{1.07E-05}} & 0.251 & \multicolumn{1}{c}{0.119408 } & 0.119408  & 2.19E-10 & 0.45  & 0.354692  & 0.354692  & 0 \\
    \textcolor[rgb]{ 1,  0,  0}{\textbf{0.006}} & \textcolor[rgb]{ 1,  0,  0}{\textbf{0.000218 }} & \textcolor[rgb]{ 1,  0,  0}{\textbf{0.000072 }} & \textcolor[rgb]{ 1,  0,  0}{\textbf{9.30E-05}} & 0.131 & 0.033781  & 0.033765  & 8.02E-06 & 0.256 & \multicolumn{1}{c}{0.123984 } & 0.123984  & 1.21E-10 & 0.46  & 0.369169  & 0.369169  & 0 \\
    \textcolor[rgb]{ 1,  0,  0}{\textbf{0.011}} & \textcolor[rgb]{ 1,  0,  0}{\textbf{0.000494 }} & \textcolor[rgb]{ 1,  0,  0}{\textbf{0.000242 }} & \textcolor[rgb]{ 1,  0,  0}{\textbf{1.48E-04}} & 0.136 & 0.036359  & 0.036347  & 5.96E-06 & 0.261 & \multicolumn{1}{c}{0.128635 } & 0.128635  & 6.61E-11 & 0.47  & 0.383882  & 0.383882  & 0 \\
    \textcolor[rgb]{ 1,  0,  0}{\textbf{0.016}} & \textcolor[rgb]{ 1,  0,  0}{\textbf{0.000842 }} & \textcolor[rgb]{ 1,  0,  0}{\textbf{0.000512 }} & \textcolor[rgb]{ 1,  0,  0}{\textbf{1.86E-04}} & 0.141 & 0.039029  & 0.039020  & 4.37E-06 & 0.266 & \multicolumn{1}{c}{0.133360 } & 0.133360  & 3.53E-11 & 0.48  & 0.398832  & 0.398832  & 0 \\
    \textcolor[rgb]{ 1,  0,  0}{\textbf{0.021}} & \textcolor[rgb]{ 1,  0,  0}{\textbf{0.001265 }} & \textcolor[rgb]{ 1,  0,  0}{\textbf{0.000882 }} & \textcolor[rgb]{ 1,  0,  0}{\textbf{2.09E-04}} & 0.146 & 0.041789  & 0.041783  & 3.17E-06 & 0.271 & \multicolumn{1}{c}{0.138160 } & 0.138160  & 1.86E-11 & 0.5   & 0.429433  & 0.429433  & 0 \\
    \textcolor[rgb]{ 1,  0,  0}{\textbf{0.026}} & \textcolor[rgb]{ 1,  0,  0}{\textbf{0.001766 }} & \textcolor[rgb]{ 1,  0,  0}{\textbf{0.001351 }} & \textcolor[rgb]{ 1,  0,  0}{\textbf{2.21E-04}} & 0.151 & 0.044639  & 0.044635  & 2.28E-06 & 0.276 & \multicolumn{1}{c}{0.143033 } & 0.143033  & 9.68E-12 & 0.55  & 0.509989  & 0.509989  & 0 \\
    \textcolor[rgb]{ 1,  0,  0}{\textbf{0.031}} & \textcolor[rgb]{ 1,  0,  0}{\textbf{0.002348 }} & \textcolor[rgb]{ 1,  0,  0}{\textbf{0.001920 }} & \textcolor[rgb]{ 1,  0,  0}{\textbf{2.24E-04}} & 0.156 & 0.047579  & 0.047575  & 1.62E-06 & 0.281 & \multicolumn{1}{c}{0.147978 } & 0.147978  & 5.04E-12 & 1.05  & 1.643087  & 1.643087  & 0 \\
    \textcolor[rgb]{ 1,  0,  0}{\textbf{0.036}} & \textcolor[rgb]{ 1,  0,  0}{\textbf{0.003014 }} & \textcolor[rgb]{ 1,  0,  0}{\textbf{0.002589 }} & \textcolor[rgb]{ 1,  0,  0}{\textbf{2.20E-04}} & 0.161 & 0.050606  & 0.050604  & 1.13E-06 & 0.286 & 0.152996  & 0.152996  & 2.55E-12 & 1.55  & 3.535109  & 3.535109  & 0 \\
    0.041 & 0.003767  & 0.003356  & 2.11E-04 & 0.166 & 0.053721  & 0.053719  & 7.83E-07 & 0.291 & 0.158086  & 0.158086  & 1.29E-12 & 2.05  & 6.631814  & 6.631814  & 0 \\
    0.046 & 0.004611  & 0.004223  & 1.98E-04 & 0.171 & 0.056923  & 0.056922  & 5.34E-07 & 0.296 & 0.163247  & 0.163247  & 6.42E-13 & 2.55  & 11.706205  & 11.706205  & 0 \\
    0.051 & 0.005548  & 0.005189  & 1.83E-04 & \textcolor[rgb]{ 1,  0,  0}{\textbf{0.176}} & \textcolor[rgb]{ 1,  0,  0}{\textbf{0.060211 }} & \textcolor[rgb]{ 1,  0,  0}{\textbf{0.060210 }} & \textcolor[rgb]{ 1,  0,  0}{\textbf{3.60E-07}} & 0.3   & 0.167427  & 0.167427  & 3.64E-13 & 3.05  & 20.042707  & 20.042707  & 0 \\
    0.056 & 0.006579  & 0.006253  & 1.66E-04 & 0.181 & 0.063584  & 0.063584  & 2.41E-07 & 0.31  & 0.178072  & 0.178072  & 8.34E-14 & 3.55  & 33.762208  & 33.762208  & 0 \\
    0.061 & 0.007707  & 0.007415  & 1.48E-04 & 0.186 & 0.067042  & 0.067042  & 1.59E-07 & 0.32  & 0.188994  & 0.188994  & 1.86E-14 & 4.05  & 56.362141  & 56.362141  & 0 \\
    0.066 & 0.008933  & 0.008675  & 1.30E-04 & 0.191 & 0.070584  & 0.070584  & 1.03E-07 & 0.33  & 0.200190  & 0.200190  & 0     & 4.55  & 93.608358  & 93.608358  & 0 \\
    0.071 & 0.010256  & 0.010032  & 1.13E-04 & 0.196 & 0.074210  & 0.074210  & 6.66E-08 & 0.34  & 0.211656  & 0.211656  & 0     & 5.05  & 155.006278  & 155.006278  & 0 \\
    0.076 & 0.011679  & 0.011487  & 9.67E-05 & 0.201 & 0.077918  & 0.077918  & 4.21E-08 & 0.35  & 0.223387  & 0.223387  & 0     & 5.55  & 256.226767  & 256.226767  & 0 \\
    0.081 & 0.013201  & 0.013038  & 8.20E-05 & 0.206 & 0.081709  & 0.081709  & 2.65E-08 & 0.36  & 0.235381  & 0.235381  & 0     & 6.05  & 423.105897  & 423.105897  & 0 \\
    0.086 & 0.014822  & 0.014685  & 6.84E-05 & 0.211 & 0.085580  & 0.085580  & 1.62E-08 & 0.37  & 0.247635  & 0.247635  & 0     & 6.55  & 698.239490  & 698.239490  & 0 \\
    0.091 & 0.016541  & 0.016429  & 5.65E-05 & 0.216 & 0.089533  & 0.089533  & 9.96E-09 & 0.38  & 0.260146  & 0.260146  & 0     & 7.05  & 1151.855685  & 1151.855685  & 0 \\
    0.096 & 0.018359  & 0.018267  & 4.61E-05 & 0.221 & 0.093566  & 0.093566  & 5.98E-09 & 0.39  & 0.272910  & 0.272910  & 0     & 7.55  & 1899.740745  & 1899.740745  & 0 \\
    0.101 & 0.020275  & 0.020201  & 3.72E-05 & 0.226 & 0.097678  & 0.097678  & 3.57E-09 & 0.4   & 0.285926  & 0.285926  & 0     & 8.05  & 3132.793687  & 3132.793687  & 0 \\
    0.106 & 0.022288  & 0.022229  & 2.97E-05 & 0.231 & 0.101869  & 0.101869  & 2.11E-09 & 0.41  & 0.299191  & 0.299191  & 0     & 8.55  & 5165.753600  & 5165.753600  & 0 \\
    0.111 & 0.024397  & 0.024350  & 2.33E-05 & 0.236 & 0.106138  & 0.106138  & 1.22E-09 & 0.42  & 0.312701  & 0.312701  & 0     & 9.05  & 8517.537393  & 8517.537393  & 0 \\
    0.116 & 0.026602  & 0.026565  & 1.82E-05 & 0.241 & 0.110484  & 0.110484  & 6.99E-10 & 0.43  & 0.326457  & 0.326457  & 0     & 9.55  & 14043.694332  & 14043.694332  & 0 \\
    0.121 & 0.028901  & 0.028873  & 1.40E-05 & 0.246 & 0.114908  & 0.114908  & 3.94E-10 & 0.44  & 0.340454  & 0.340454  & 0     & 10    & 22025.465568  & 22025.465568  & 0 \\

    \bottomrule[1.5pt]

    \multirow{2}{*}{$\vartheta$} &
    \multicolumn{3}{c}{$R(\gamma)=0.5$} &\multicolumn{3}{c}{$R(\gamma)=1$} &\multicolumn{3}{c}{$R(\gamma)=1.5$} &\multicolumn{3}{c}{$R(\gamma)=2$} &\multicolumn{3}{c}{$R(\gamma)=4$}\cr
    \cmidrule[0.5pt](lr){2-4} \cmidrule[0.5pt](lr){5-7} \cmidrule[0.5pt](lr){8-10} \cmidrule[0.5pt](lr){11-13} \cmidrule[0.5pt] (lr){14-16}
    &$\nu_{3}$ &Exact &$\varpi\left(\nu_{3}\right)$ &$\nu_{3}$ &Exact &$\varpi\left(\nu_{3}\right)$ &$\nu_{3}$ &Exact &$\varpi\left(\nu_{3}\right)$ &$\nu_{3}$ &Exact &$\varpi\left(\nu_{3}\right)$ &$\nu_{3}$ &Exact &$\varpi\left(\nu_{3}\right)$ \cr
    \midrule[0.8pt]
    0.01  & 0.661943     & 0.661943     & 0 & 1.743713     & 1.743713     & 0 & 3.525612     & 3.525612     & 0 & 6.462644      & 6.462644      & 0 & 54.146780      & 54.146780      & 0 \\
    0.06  & 0.731489     & 0.731489     & 0 & 1.875582     & 1.875582     & 0 & 3.752433     & 3.752433     & 0 & 6.842128      & 6.842128      & 0 & 56.973794      & 56.973794      & 0 \\
    0.11  & 0.806910     & 0.806910     & 0 & 2.015530     & 2.015530     & 0 & 3.991668     & 3.991668     & 0 & 7.241540      & 7.241540      & 0 & 59.945815      & 59.945815      & 0 \\
    0.16  & 0.888390     & 0.888390     & 0 & 2.163876     & 2.163876     & 0 & 4.243891     & 4.243891     & 0 & 7.661866      & 7.661866      & 0 & 63.070274      & 63.070274      & 0 \\
    0.21  & 0.976102     & 0.976102     & 0 & 2.320958     & 2.320958     & 0 & 4.509711     & 4.509711     & 0 & 8.104141      & 8.104141      & 0 & 66.354981      & 66.354981      & 0 \\
    0.26  & 1.070221     & 1.070221     & 0 & 2.487133     & 2.487133     & 0 & 4.789770     & 4.789770     & 0 & 8.569457      & 8.569457      & 0 & 69.808147      & 69.808147      & 0 \\
    0.31  & 1.170924     & 1.170924     & 0 & 2.662782     & 2.662782     & 0 & 5.084745     & 5.084745     & 0 & 9.058965      & 9.058965      & 0 & 73.438407      & 73.438407      & 0 \\
    0.36  & 1.278397     & 1.278397     & 0 & 2.848307     & 2.848307     & 0 & 5.395354     & 5.395354     & 0 & 9.573876      & 9.573876      & 0 & 77.254834      & 77.254834      & 0 \\
    0.41  & 1.392837     & 1.392837     & 0 & 3.044136     & 3.044136     & 0 & 5.722353     & 5.722353     & 0 & 10.115466     & 10.115466     & 0 & 81.266971      & 81.266971      & 0 \\
    0.46  & 1.514458     & 1.514458     & 0 & 3.250725     & 3.250725     & 0 & 6.066539     & 6.066539     & 0 & 10.685075     & 10.685075     & 0 & 85.484850      & 85.484850      & 0 \\
    0.50  & 1.617081     & 1.617081     & 0 & 3.424067     & 3.424067     & 0 & 6.354827     & 6.354827     & 0 & 11.161886     & 11.161886     & 0 & 89.014354      & 89.014354      & 0 \\
    1.00  & 3.364012     & 3.364012     & 0 & 6.320107     & 6.320107     & 0 & 11.141172    & 11.141172    & 0 & 19.060581     & 19.060581     & 0 & 147.409790     & 147.409790     & 0 \\
    1.50  & 6.284876     & 6.284876     & 0 & 11.120350    & 11.120350    & 0 & 19.048069    & 19.048069    & 0 & 32.092775     & 32.092775     & 0 & 243.688867     & 243.688867     & 0 \\
    2.00  & 11.099420    & 11.099420    & 0 & 19.035532    & 19.035532    & 0 & 32.085206    & 32.085206    & 0 & 53.579824     & 53.579824     & 0 & 402.426315     & 402.426315     & 0 \\
    2.50  & 19.022972    & 19.022972    & 0 & 32.077632    & 32.077632    & 0 & 53.575239    & 53.575239    & 0 & 89.003241     & 89.003241     & 0 & 664.139754     & 664.139754     & 0 \\
    3.00  & 32.070052    & 32.070052    & 0 & 53.570653    & 53.570653    & 0 & 89.000463    & 89.000463    & 0 & 147.403051    & 147.403051    & 0 & 1095.631791    & 1095.631791    & 0 \\
    3.50  & 53.566067    & 53.566067    & 0 & 88.997684    & 88.997684    & 0 & 147.401366   & 147.401366   & 0 & 243.684780    & 243.684780    & 0 & 1807.041447    & 1807.041447    & 0 \\
    4.00  & 88.994904    & 88.994904    & 0 & 147.399681   & 147.399681   & 0 & 243.683758   & 243.683758   & 0 & 402.423836    & 402.423836    & 0 & 2979.957316    & 2979.957316    & 0 \\
    4.50  & 147.397996   & 147.397996   & 0 & 243.682736   & 243.682736   & 0 & 402.423216   & 402.423216   & 0 & 664.138250    & 664.138250    & 0 & 4913.768382    & 4913.768382    & 0 \\
    5.00  & 243.681715   & 243.681715   & 0 & 402.422596   & 402.422596   & 0 & 664.137874   & 664.137874   & 0 & 1095.630879   & 1095.630879   & 0 & 8102.083619    & 8102.083619    & 0 \\
    5.50  & 402.421977   & 402.421977   & 0 & 664.137499   & 664.137499   & 0 & 1095.630651  & 1095.630651  & 0 & 1807.040893   & 1807.040893   & 0 & 13358.726624   & 13358.726624   & 0 \\
    6.00  & 664.137123   & 664.137123   & 0 & 1095.630423  & 1095.630423  & 0 & 1807.040755  & 1807.040755  & 0 & 2979.956981   & 2979.956981   & 0 & 22025.465659   & 22025.465659   & 0 \\
    6.50  & 1095.630195  & 1095.630195  & 0 & 1807.040617  & 1807.040617  & 0 & 2979.956897  & 2979.956897  & 0 & 4913.768179   & 4913.768179   & 0 & 36314.502585   & 36314.502585   & 0 \\
    7.00  & 1807.040479  & 1807.040479  & 0 & 2979.956813  & 2979.956813  & 0 & 4913.768128  & 4913.768128  & 0 & 8102.083496   & 8102.083496   & 0 & 59873.141657   & 59873.141657   & 0 \\
    7.50  & 2979.956729  & 2979.956729  & 0 & 4913.768077  & 4913.768077  & 0 & 8102.083465  & 8102.083465  & 0 & 13358.726549  & 13358.726549  & 0 & 98714.770973   & 98714.770973   & 0 \\
    8.00  & 4913.768026  & 4913.768026  & 0 & 8102.083434  & 8102.083434  & 0 & 13358.726530 & 13358.726530 & 0 & 22025.465613  & 22025.465613  & 0 & 162753.791394  & 162753.791394  & 0 \\
    8.50  & 8102.083403  & 8102.083403  & 0 & 13358.726512 & 13358.726512 & 0 & 22025.465602 & 22025.465602 & 0 & 36314.502557  & 36314.502557  & 0 & 268336.286505  & 268336.286505  & 0 \\
    9.00  & 13358.726493 & 13358.726493 & 0 & 22025.465591 & 22025.465591 & 0 & 36314.502550 & 36314.502550 & 0 & 59873.141640  & 59873.141640  & 0 & 442412.391999  & 442412.391999  & 0 \\
    9.50  & 22025.465579 & 22025.465579 & 0 & 36314.502543 & 36314.502543 & 0 & 59873.141636 & 59873.141636 & 0 & 98714.770963  & 98714.770963  & 0 & 729415.369841  & 729415.369841  & 0 \\
    10.00 & 36314.502537 & 36314.502537 & 0 & 59873.141632 & 59873.141632 & 0 & 98714.770960 & 98714.770960 & 0 & 162753.791388 & 162753.791388 & 0 & 1202603.284161 & 1202603.284161 & 0 \\
    \bottomrule[1.5pt]
    \end{tabular}
    \end{threeparttable}}
\end{table*}

\begin{remark}
In Fig.~\ref{GammaAnalysis}, we describe the properties of $\nu_{3}$ with different values of decoding error probability, i.e., $\epsilon$, and blocklength, i.e., $n$.  From Fig.~\ref{GammaAnalysis}, it's easy to find that with a small value of $n$, the decoding error probability has a significant impact on $\nu_{3}$. As expected, a lower decoding error probability yields a higher $\nu_{3}$. This is because achieving a lower decoding error probability naturally requires higher SNR or (SINR) for transmitting a given data bits $D$. Moreover, the value of blocklength also greatly affects the value of $\nu_{3}$ which decreases a lot with the increasing of $n$. This is also not difficult to understand that increasing the blocklength $n$, i.e. the number of channels used $n$, means that the number of bits transmitting on each channel decreases for transmitting a given data bits $D$. Consequently, it naturally requires a lower value of SNR (SINR) for transmitting a given data bits $D$ and achieving a certain decoding error probability.
\begin{figure}[t]
\renewcommand{\captionfont}{\footnotesize}
	\centering
	\includegraphics[width=0.5\columnwidth,keepaspectratio]{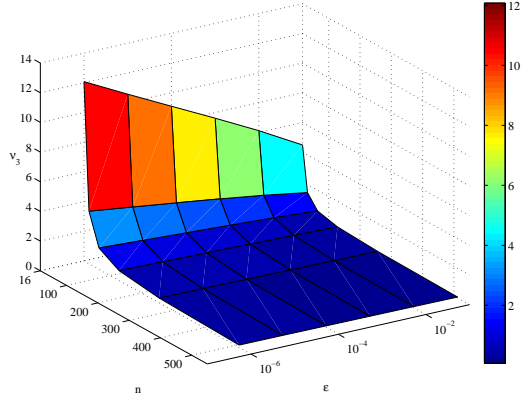}\\
	\caption{Illustration of $\nu_{3}$.}
	\label{GammaAnalysis}
\end{figure}
\end{remark}
\begin{theorem}\label{URLLCTheorem03}
Given $\vartheta>0$, function $R\left(\gamma\right)$ has an inflection point $\nu_{4}$, i.e., the concave-convex property of function $R\left(\gamma\right)$ changes in $0<\gamma<\infty$.
\end{theorem}
\begin{IEEEproof}
Let $g\left(\gamma\right)=\frac{3\left(1+\gamma\right)^{2}-2}{\left(1+\gamma\right)\left(\left(1+\gamma\right)^{2}-1\right)^{\frac{3}{2}}}$. The first derivation of $g\left(\gamma\right)$ is calculated as
\begin{equation}\label{URLLC25}
g'\left(\gamma\right)=\frac{-6\left(1+\gamma\right)^{4}+\left(1+\gamma\right)^{2}-2}{\left(1+\gamma\right)^{2}\left(\left(1+\gamma\right)^{2}-1\right)^{\frac{5}{2}}}
\end{equation}
Due to $\gamma>0$, we further have
\begin{equation}\label{URLLC26}
g'\left(\gamma\right)<\frac{-5\left(1+\gamma\right)^{4}-2}{\left(1+\gamma\right)^{2}\left(\left(1+\gamma\right)^{2}-1\right)^{\frac{5}{2}}}<0
\end{equation}
implying function $g\left(\gamma\right)$ is a monotonic decreasing function with respect to $\gamma>0$. Therefore, for given $\forall \epsilon$ and $\forall n$, there is a $\nu_{4}$ that makes $\vartheta g\left(\nu_{4}\right)=1$ hold. We further have $R''\left(\gamma\right)>0$ for $0<\gamma<\nu_{4}$ and $R''\left(\gamma\right)<0$ for $\nu_{4}<\gamma$. This implies that $\nu_{4}$ is the dividing point between concave and convex curve of function $R\left(\gamma\right)$, i.e., $\nu_{4}$ is the inflection point of function $R\left(\gamma\right)$.
\end{IEEEproof}
\begin{remark}
Fig.~\ref{ConcaveConvexAnalysis} illustrates the curve of functions $R\left(\gamma\right)$ and $g\left(\gamma\right)$ for different parameters. The numerical results verify the conclusions obtained in the previous analysis for the proofs of Theorem~\ref{URLLCTheorem01} and~\ref{URLLCTheorem03}. Namely, function $R\left(\gamma\right)$ is a non-convex and non-concave function with respect to $\gamma$. Furthermore, the inflection point of concave-convex property of function $R\left(\gamma\right)$ seems to be disappeared as the value of $\vartheta$ decreases. This is because once the value of $\vartheta$ is extreme small, the property of function $R\left(\gamma\right)$ is mainly determined by function $C\left(\gamma\right)$, which is concave with respect to $\gamma$. The subfigure on the right hand illustrates the monotonic decreasing of function $g\left(\gamma\right)$ with respect to $\gamma$. In addition, unfortunately, the analytical expression of inflection point $\nu_{4}$ cannot be given out in Theorem~\ref{URLLCTheorem03}. This also brings us a barrier to judge the convexity and concavity of function $R\left(\gamma\right)$ in the range $\gamma>\nu_{3}$. Consequently, these observations make problems~\eqref{URLLC05},~\eqref{URLLC06}, and~\eqref{URLLC07} be non-convex and non-concave. Generally speaking, it is difficult to obtain their global optimal solutions, even their local optimal solutions. In other words, directly solving problems~\eqref{URLLC05},~\eqref{URLLC06}, and~\eqref{URLLC07} is challenging.
\end{remark}
\begin{figure}[t]
\renewcommand{\captionfont}{\footnotesize}
	\centering
	\includegraphics[width=0.6\columnwidth,keepaspectratio]{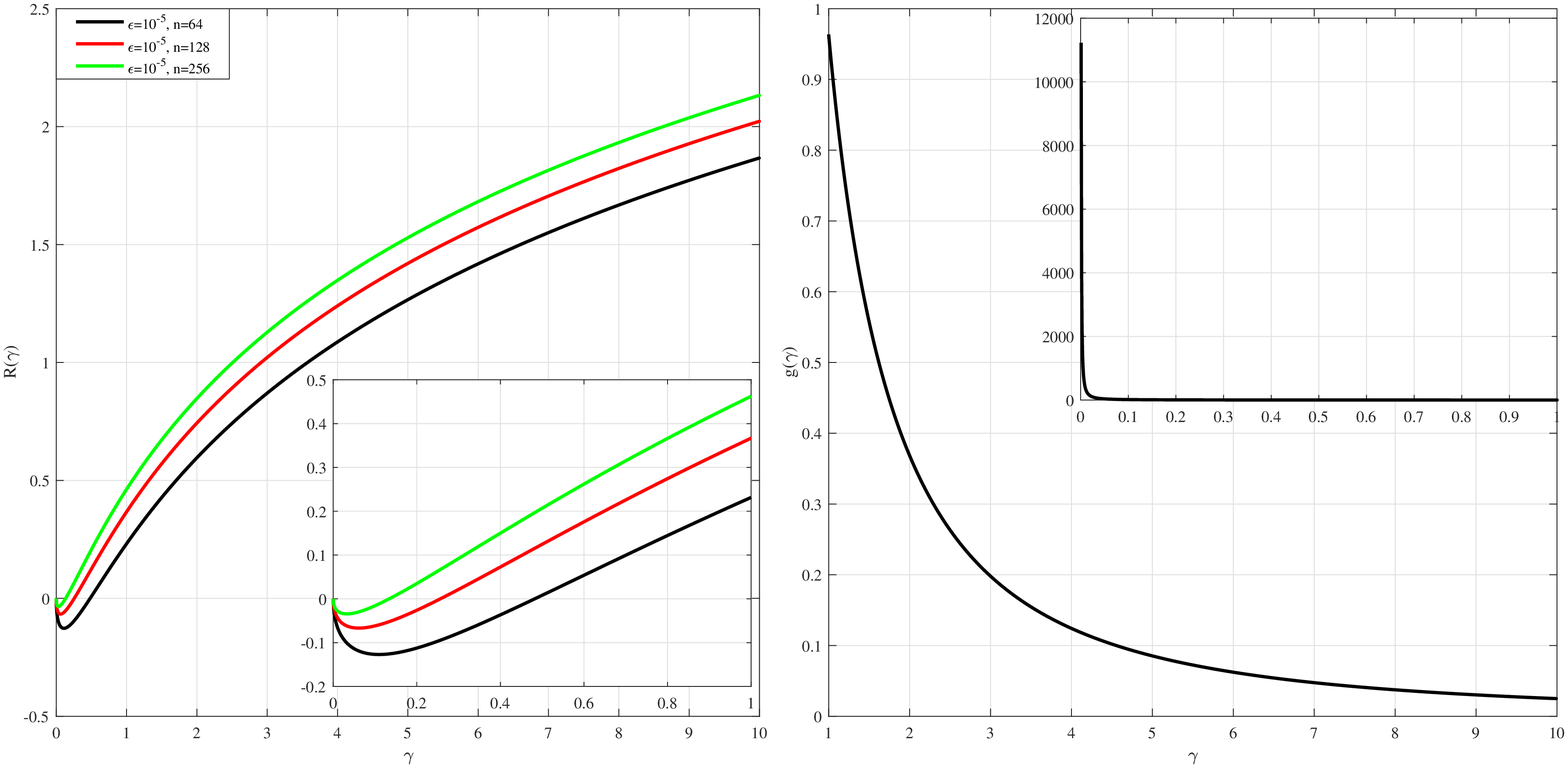}\\
	\caption{Illustration of functions $R\left(\gamma\right)$ and $g\left(\gamma\right)$.}
	\label{ConcaveConvexAnalysis}
\end{figure}

\section{ \label {beamforming design}Beamforming Design}

In this section, we focus on designing an efficient optimization algorithm to address problems~\eqref{URLLC05},~\eqref{URLLC06}, and~\eqref{URLLC07} via some basic mathematical operations and successive convex approximation method~\cite{CLNguyen2015}. It is well known that the uplink-downlink duality theory has shown that the uplink and downlink communications have the same SINR region for the downlink multiuser communication systems subject to the total power constraint~\cite{TVTSchubert2004,TCOMHe2015}. In the sequel, we revisit to use the uplink-downlink duality theory to address problems~\eqref{URLLC05},~\eqref{URLLC06}, and~\eqref{URLLC07} instead of directly solving them. For ease of notation, we let $q_{k}$ be the transmitting power of the uplink $k$-th user and firstly give out the following Lemmas.
\begin{lemma}\label{URLLCLemma01}
Given the optimal solution $\left\{\mathbf{w}_{k}^{*}, q_{k}^{*}\right\}$ for the uplink SRMax problem, i.e.,
\begin{subequations}\label{URLLC27}
\begin{align}
&\max_{\left\{\mathbf{w}_{k},q_{k}\right\}}~\sum\limits_{k\in\mathcal{K}}\alpha_{k}\overleftarrow{R}\left(\overleftarrow{\gamma}_{k}\right),\label{URLLC27a}\\
\mathrm{s.t.}~&\sum\limits_{k\in\mathcal{K}}q_{k}\leq P, \left\|\mathbf{w}_{k}\right\|_{2}=1, \forall k\in\mathcal{K}, \label{URLLC27b}\\
&\frac{D}{n}\ln\left(2\right)\leq \overleftarrow{R}\left(\overleftarrow{\gamma}_{k}\right), \forall k\in\mathcal{K}. \label{URLLC27c}
\end{align}
\end{subequations}
where $\overleftarrow{R}\left(\overleftarrow{\gamma}_{k}\right)=C\left(\overleftarrow{\gamma}_{k}\right)-\vartheta
\sqrt{V\left(\overleftarrow{\gamma}_{k}\right)}$ and $\overleftarrow{\gamma}_{k}$ is given by
\begin{equation}\label{URLLC28}
\overleftarrow{\gamma}_{k}=\frac{q_{k}\left|\overline{\mathbf{h}}_{k}^{H}\mathbf{w}_{k}\right|^{2}}
{\sum\limits_{l\in\mathcal{K}, l\neq k}q_{l}\left|\overline{\mathbf{h}}_{l}^{H}\mathbf{w}_{k}\right|^{2}+1}.
\end{equation}

Then, $\left\{\mathbf{w}_{k}^{*}, p_{k}^{*}\right\}$ is the optimal solution to the downlink SRMax problem~\eqref{URLLC05}, where $p_{k}^{*}$ is the $k$-th element of vector $\mathbf{p}^{*}=\left[p_{1}^{*},\cdots,p_{K}^{*}\right]^{T}$, which is calculated as:
\begin{equation}\label{URLLC29}
\mathbf{p}^{*}=\mathbf{\Psi}^{-1}\mathbf{1},
\end{equation}
with matrix $\mathbf{\Psi}$  satisfying
\begin{equation}\label{URLLC30}
\left[\mathbf{\Psi}\right]_{k,l}=\left\{
\begin{aligned}
\frac{\left|\overline{\mathbf{h}}_{k}^{H}\mathbf{w}_{k}^{*}\right|^{2}}{\overleftarrow{\gamma}_{k}^{*}}, k=l \\
-\left|\overline{\mathbf{h}}_{k}^{H}\mathbf{w}_{l}^{*}\right|^{2}, k\neq l,
\end{aligned}
\right.
\end{equation}
where $\overleftarrow{\gamma}_{k}^{*}$ is calculated with $\left\{\mathbf{w}_{k}^{*}, q_{k}^{*}\right\}$ via~\eqref{URLLC28} and $\mathbf{1}$ is an all one vector.
\end{lemma}

\begin{lemma}\label{URLLCLemma02}
Given the optimal solution $\left\{\mathbf{w}_{k}^{*}, q_{k}^{*}\right\}$ for the uplink EEMax problem, i.e.,
\begin{subequations}\label{URLLC31}
\begin{align}
&\max_{\left\{\mathbf{w}_{k},q_{k}\right\}}~\frac{\sum\limits_{k\in\mathcal{K}}\alpha_{k}\overleftarrow{R}\left(\overleftarrow{\gamma}_{k}\right)}
{\eta\sum\limits_{k\in\mathcal{K}}q_{k}+N_{\mathrm{t}}P_{\mathrm{c}}+P_{0}},\label{URLLC31a}\\
\mathrm{s.t.}~&\sum\limits_{k\in\mathcal{K}}q_{k}\leq P, \left\|\mathbf{w}_{k}\right\|_{2}=1, \forall k\in\mathcal{K}, \label{URLLC31b}\\
&\frac{D}{n}\ln\left(2\right)\leq \overleftarrow{R}\left(\overleftarrow{\gamma}_{k}\right), \forall k\in\mathcal{K}. \label{URLLC31c}
\end{align}
\end{subequations}
Then, $\left\{\mathbf{w}_{k}^{*}, p_{k}^{*}\right\}$ is the optimal solution of the downlink EEMax problem~\eqref{URLLC06}, where $p_{k}^{*}$ is the $k$-th element of vector $\mathbf{p}^{*}=\left[p_{1}^{*},\cdots,p_{K}^{*}\right]^{T}$, which is calculated as~\eqref{URLLC29}.
\end{lemma}

\begin{lemma}\label{URLLCLemma03}
Given the optimal solution $\left\{\mathbf{w}_{k}^{*}, q_{k}^{*}\right\}$ for the uplink fairness problem, i.e.,
\begin{subequations}\label{URLLC32}
\begin{align}
&\max_{\left\{\mathbf{w}_{k},q_{k}\right\}}\min\limits_{k\in\mathcal{K}}~R\left(\overleftarrow{\gamma}_{k}\right),\label{URLLC32a}\\
\mathrm{s.t.}~&\sum\limits_{k\in\mathcal{K}}q_{k}\leq P, \left\|\mathbf{w}_{k}\right\|_{2}^{2}=1, \forall k\in\mathcal{K}, \label{URLLC32b}\\
&\frac{D}{n}\ln\left(2\right)\leq R\left(\overleftarrow{\gamma}_{k}\right), \forall k\in\mathcal{K}. \label{URLLC32c}
\end{align}
\end{subequations}
Then, $\left\{\mathbf{w}_{k}^{*}, p_{k}^{*}\right\}$ is the optimal solution of the downlink MaxMin problem~\eqref{URLLC07}, where $p_{k}^{*}$ is the $k$-th element of vector $\mathbf{p}^{*}=\left[p_{1}^{*},\cdots,p_{K}^{*}\right]^{T}$, which is calculated as~\eqref{URLLC29}.
\end{lemma}

From the expression of the uplink SINR $\overleftarrow{\gamma}_{k}$ in~\eqref{URLLC28} and the descriptions of problems~\eqref{URLLC27},~\eqref{URLLC31}, and~\eqref{URLLC32}, it is not difficult to find that,  given the uplink transmitting power $\left\{q_{k}\right\}$, solving problems~\eqref{URLLC27},~\eqref{URLLC31}, and~\eqref{URLLC32} can resort to maximize $\overleftarrow{R}\left(\overleftarrow{\gamma}_{k}\right)$ with respect to $\left\{\mathbf{w}_{k}\right\}$. In the sequel, we focus on addressing problems~\eqref{URLLC27},~\eqref{URLLC31}, and~\eqref{URLLC32} rather than directly solving problems~\eqref{URLLC05},~\eqref{URLLC06}, and~\eqref{URLLC07} via two-step. In particular, we adopt the block coordinate descent method to design an iterative optimization algorithm to address them. Given the uplink transmitting power $\left\{q_{k}\right\}$, the optimal solution of $\left\{\mathbf{w}_{k}\right\}$ for maximizing $\overleftarrow{R}\left(\overleftarrow{\gamma}_{k}\right)$ is the minimum mean square error receiver, i.e.,
\begin{equation}\label{URLLC33}
\mathbf{w}_{k}^{*}=\frac{\left(\mathbf{I}_{N_{\mathrm{t}}}+\sum\limits_{k\in\mathcal{K}}q_{k}\overline{\mathbf{h}}_{k}\overline{\mathbf{h}}_{k}^{H}\right)^{-1}\overline{\mathbf{h}}_{k}}
{\left\|\left(\mathbf{I}_{N_{\mathrm{t}}}+\sum\limits_{k\in\mathcal{K}}q_{k}\overline{\mathbf{h}}_{k}\overline{\mathbf{h}}_{k}^{H}\right)^{-1}\overline{\mathbf{h}}_{k}\right\|}
\end{equation}
where $\mathbf{I}_{N_{\mathrm{t}}}$ denotes $N_{\mathrm{t}}$-by-$N_{\mathrm{t}}$ identity matrix.

\subsection{For SR Maximization}

Recalling the conclusions obtained in Theorems~\ref{URLLCTheorem01} and~\ref{URLLCTheorem02}, i.e., the monotonic increasing property of function $\widetilde{R}\left(\overleftarrow{\gamma}_{k}\right)$ with respect to $\overleftarrow{\gamma}_{k}$ in the range $\nu_{3}\leq\overleftarrow{\gamma}_{k}\leq\tilde{\gamma}_{k}$,  $\frac{D}{n}\ln\left(2\right)\leq \overleftarrow{R}\left(\overleftarrow{\gamma}_{k}\right)$ can be rewritten as $\nu_{3}\leq\overleftarrow{\gamma}_{k}$, $\forall k\in\mathcal{K}$. Accordingly, given the beamforming vector $\left\{\mathbf{w}_{k}\right\}$, problem~\eqref{URLLC27} is reformulated as
\begin{subequations}\label{URLLC34}
\begin{align}
&\max_{\left\{q_{k}\right\}}~\sum\limits_{k\in\mathcal{K}}\alpha_{k}\overleftarrow{R}\left(\overleftarrow{\gamma}_{k}\right),\label{URLLC34a}\\
\mathrm{s.t.}~&\sum\limits_{k\in\mathcal{K}}q_{k}\leq P,~\nu_{3}\leq\overleftarrow{\gamma}_{k},~\forall k\in\mathcal{K}. \label{URLLC34b}
\end{align}
\end{subequations}
Before proceeding, to obtain a tractable form of problem~\eqref{URLLC34}, we equivalently rewrite it as problem~\eqref{URLLC35} by introducing auxiliary variables $\varphi_{k}$, $\psi_{k}$, $\phi_{k}$, and $\theta_{k}$, $\forall k\in\mathcal{K}$.
\begin{subequations}\label{URLLC35}
\begin{align}
&\max~\sum\limits_{k\in\mathcal{K}}\alpha_{k}\widetilde{R}\left(\varphi_{k},\theta_{k}\right),\label{URLLC35a}\\
\mathrm{s.t.}~&\nu_{3}\leq\varphi_{k}, \forall k\in\mathcal{K},\label{URLLC35b}\\
&\varphi_{k}\leq\overleftarrow{\gamma}_{k}, \forall k\in\mathcal{K},\label{URLLC35c}\\
&\phi_{k}\leq\tilde{\gamma}_{k}, \forall k\in\mathcal{K},\label{URLLC35d}\\
&\overleftarrow{\gamma}_{k}\leq\phi_{k}, \forall k\in\mathcal{K},\label{URLLC35e}\\
&\widetilde{V}\left(\phi_{k}\right)\leq\psi_{k}, \forall k\in\mathcal{K},\label{URLLC35f}\\
&\sqrt{\psi_{k}}\leq \theta_{k}, \forall k\in\mathcal{K}, \label{URLLC35g}\\
&\widetilde{V}\left(\nu_{3}\right)\leq\psi_{k}\leq \widetilde{V}\left(\tilde{\gamma}_{k}\right), \forall k\in\mathcal{K},\label{URLLC35h}\\
&\sum\limits_{k\in\mathcal{K}}q_{k}\leq P.\label{URLLC35i}
\end{align}
\end{subequations}
In problem~\eqref{URLLC35}, the optimization variables are $q_{k}$, $\varphi_{k}$, $\psi_{k}$, $\phi_{k}$, and $\theta_{k}$, $\forall k\in\mathcal{K}$. $\widetilde{R}\left(\varphi_{k},\theta_{k}\right)=\ln\left(1+\varphi_{k}\right)-\vartheta\theta_{k}$ and $\widetilde{V}\left(\phi_{k}\right)$ is given by
\begin{equation}\label{URLLC36}
\widetilde{V}\left(\phi_{k}\right)=1-\frac{1}{\left(1+\phi_{k}\right)^{2}}.
\end{equation}
At the optimal point of problem~\eqref{URLLC35}, inequality constraints~\eqref{URLLC35c} and~\eqref{URLLC35e}-\eqref{URLLC35g} are activated. It is not difficult to see that problem~\eqref{URLLC35} is non-convex due to the non-convex constraints~\eqref{URLLC35c} and~\eqref{URLLC35e}-\eqref{URLLC35g}. Constraints~\eqref{URLLC35c} and~\eqref{URLLC35e} can be equivalently rewritten as
\begin{subequations}\label{URLLC37}
\begin{align}
&\sum\limits_{l\neq k}a_{k,l}
\left|\overline{\mathbf{h}}_{l}^{H}\mathbf{w}_{k}\right|^{2}-q_{k}
\left|\overline{\mathbf{h}}_{k}^{H}\mathbf{w}_{k}\right|^{2}+\varphi_{k}\leq 0,\label{URLLC37a}\\
&q_{k}\left|\overline{\mathbf{h}}_{k}^{H}\mathbf{w}_{k}\right|^{2}-\sum\limits_{l\neq k}b_{k,l}
\left|\overline{\mathbf{h}}_{l}^{H}\mathbf{w}_{k}\right|^{2}-\phi_{k}\leq 0,\label{URLLC37b}
\end{align}
\end{subequations}
where $a_{k,l}=\varphi_{k}q_{l}$, $b_{k,l}=\phi_{k}q_{l}$, $\forall k,l\in\mathcal{K}$, and $k\neq l$. Constraint~\eqref{URLLC35f} can be equivalently rewritten as:
\begin{equation}\label{URLLC38}
\left(1+\phi_{k}\right)^{2}-\frac{1}{1-\psi_{k}}\leq 0.
\end{equation}
Exploiting the convexity of function $\frac{1}{1-\psi_{k}}$, the concavity of function $\sqrt{\psi_{k}}$, and the first-order Taylor series expansion, we can obtain their low boundary approximations as follows
\begin{subequations}\label{URLLC39}
\begin{align}
\frac{1}{1-\psi_{k}}&\geq\overline{\psi}\left(\psi_{k}\right)\triangleq\frac{1-2\psi_{k}^{\left(t\right)}}
{\left(1-\psi_{k}^{\left(t\right)}\right)^{2}}
+\frac{\psi_{k}}{\left(1-\psi_{k}^{\left(t\right)}\right)^{2}},\label{URLLC39a}\\
\sqrt{\psi_{k}}&\leq\mu\left(\psi_{k}\right)\triangleq\frac{1}{2}\sqrt{\psi_{k}^{\left(t\right)}}+\frac{\psi_{k}}
{2\sqrt{\psi_{k}^{\left(t\right)}}}.\label{URLLC39b}
\end{align}
\end{subequations}
In~\eqref{URLLC39}, $t$ denotes the $t$-th iteration, $\varphi_{k}^{\left(t\right)}$ and $\psi_{k}^{\left(t\right)}$ represent the solutions obtained at the $t$-th iteration for variables $\varphi_{k}$ and $\psi_{k}$, respectively. Thus, problem~\eqref{URLLC35} can be reformulated as 
\begin{subequations}\label{URLLC40}
\begin{align}
&\max~\sum\limits_{k\in\mathcal{K}}\alpha_{k}\widetilde{R}\left(\varphi_{k},\theta_{k}\right),\label{URLLC40a}\\
\mathrm{s.t.}~&\eqref{URLLC35b},~\eqref{URLLC35d},~\eqref{URLLC35h},~\eqref{URLLC35i},~\eqref{URLLC37},\label{URLLC40b}\\
&\left(1+\phi_{k}\right)^{2}-\overline{\psi}\left(\psi_{k}\right)\leq 0,\forall k\in\mathcal{K},\label{URLLC40c}\\
&\mu\left(\psi_{k}\right)\leq \theta_{k}, \forall k\in\mathcal{K},\label{URLLC40d}\\
&a_{k,l}=\varphi_{k}q_{l},~b_{k,l}=\phi_{k}q_{l}, \forall k,l\in\mathcal{K}, k\neq l.\label{URLLC40e}
\end{align}
\end{subequations}
In problem~\eqref{URLLC40}, the optimization variables are $q_{k}$, $\varphi_{k}$, $\psi_{k}$, $\phi_{k}$, $\theta_{k}$, $a_{k,l}$, and $b_{k,l}$, $\forall k, l\in\mathcal{K}$, $k\neq l$. Note that the bilinear form constraints in~\eqref{URLLC40e} are non-convex. To overcome these challenges, we rewrite problem~\eqref{URLLC40} into~\eqref{URLLC41} via using McCormick envelopes~\cite{CCECastro2015}, where $\tilde{q}_{k}$ be the minimum power for achieving $\nu_{3}$.
\begin{subequations}\label{URLLC41}
\begin{align}
&\max~\sum\limits_{k\in\mathcal{K}}\alpha_{k}\left(
\ln\left(1+\varphi_{k}\right)-\vartheta\theta_{k}\right),\label{URLLC41a}\\
\mathrm{s.t.}~&\eqref{URLLC40b},~\eqref{URLLC40c},~\eqref{URLLC40d},\label{URLLC41b}\\
&\nu_{3}q_{l}+\tilde{q}_{l}\varphi_{k}-a_{k,l}\leq \nu_{3}\tilde{q}_{l}, \forall k,l\in\mathcal{K}, k\neq l,\label{URLLC41c}\\
&\tilde{\gamma}_{k}q_{l}+P\varphi_{k}-a_{k,l}\leq P\tilde{\gamma}_{k}, \forall k,l\in\mathcal{K}, k\neq l,\label{URLLC41d}\\
&a_{k,l}-\tilde{\gamma}_{k}q_{l}-\tilde{q}_{l}\varphi_{k}\leq-\tilde{\gamma}_{k} \tilde{q}_{l}, \forall k,l\in\mathcal{K}, k\neq l,\label{URLLC41e}\\
&a_{k,l}-P\varphi_{k}-\nu_{3}q_{l}\leq-P\nu_{3}, \forall k,l\in\mathcal{K}, k\neq l,\label{URLLC41f}\\
&\nu_{3}q_{l}+\tilde{q}_{l}\phi_{k}-b_{k,l}\leq \tilde{q}_{l}\nu_{3}, \forall k,l\in\mathcal{K}, k\neq l,\label{URLLC41g}\\
&\tilde{\gamma}_{k}q_{l}+P\phi_{k}-b_{k,l}\leq P\tilde{\gamma}_{k}, \forall k,l\in\mathcal{K}, k\neq l,\label{URLLC41h}\\
&b_{k,l}-\tilde{\gamma}_{k}q_{l}-\tilde{q}_{l}\phi_{k}\leq -\tilde{q}_{l}\tilde{\gamma}_{k}, \forall k,l\in\mathcal{K}, k\neq l,\label{URLLC41i}\\
&b_{k,l}-P\phi_{k}-\nu_{3}q_{l}\leq -P\nu_{3}, \forall k,l\in\mathcal{K}, k\neq l,\label{URLLC41j}
\end{align}
\end{subequations}
In problem~\eqref{URLLC41}, the optimization variables are $q_{k}$, $\varphi_{k}$, $\psi_{k}$, $\phi_{k}$, $\theta_{k}$, $a_{k,l}$, and $b_{k,l}$, $\forall k, l\in\mathcal{K}$, $k\neq l$. Problem~\eqref{URLLC41} is convex and can be easily solved using the classical optimization methods, e.g., primal-dual interior-point methods, with complexity $\mathcal{O}\left(\left(K^{2}+3K\right)^{3.5}+N_{\mathrm{t}}^{2.7}\right)$~\cite{ComPotra2000,MatrixComplexity}. The detailed steps of solving problem~\eqref{URLLC21} is outlined in Algorithm~\ref{URLLCA01} where $\xi^{\left(\tau\right)}$ and $\varsigma^{\left(t\right)}$ denote the objective values of problem~\eqref{URLLC27} and problem~\eqref{URLLC41} at the $\tau$-th and $t$-th iterations, respectively. A nondecreasing sequence of objective values $\varsigma^{\left(t\right)}$ can be generated via running Algorithm~\ref{URLLCA01} with similar analysis as in~\cite{CLNguyen2015}. Meanwhile, due to the limited allowable transmitting power, the problem has an upper bound. This implies that the convergence of Algorithm~\ref{URLLCA01} can be guaranteed by the monotonic boundary theorem~\cite{Bibby1974}.
\begin{algorithm}[htp]
\caption{Solution of problem~\eqref{URLLC27}}\label{URLLCA01}
\begin{algorithmic}[1]
\STATE Initialize beamforming vector $\mathbf{w}_{k}^{\left(0\right)}$ and $p_{k}^{\left(0\right)}$, $\forall k\in\mathcal{K}$, such that constraint~\eqref{URLLC05b} and~\eqref{URLLC05c} are satisfied. \label{URLLCA0101}
\STATE Compute $\gamma_{k}$ with $\mathbf{w}_{k}^{\left(0\right)}$ and $p_{k}^{\left(0\right)}$ to obtain $\bar{\gamma}_{k}$. Compute $\tilde{q}_{k}$ that is the $k$-th element of vector $\tilde{\mathbf{q}}=\left[\tilde{q}_{1},\cdots,\tilde{q}_{K}\right]^{T}$, which is calculated as: \label{URLLCA0102}
\begin{equation}\label{URLLC42}
\tilde{\mathbf{q}}=\mathbf{\Phi}^{-1}\mathbf{1}.
\end{equation}
where matrix $\mathbf{\Phi}$ is given by
\begin{equation}\label{URLLC43}
\left[\mathbf{\Phi}\right]_{k,l}=\left\{
\begin{aligned}
\frac{\left|\overline{\mathbf{h}}_{k}^{H}\mathbf{w}_{k}^{\left(0\right)}\right|^{2}}{\bar{\gamma}_{k}}, k=l \\
-\left|\overline{\mathbf{h}}_{l}^{H}\mathbf{w}_{k}^{\left(0\right)}\right|^{2}, k\neq l.
\end{aligned}
\right.
\end{equation}
\STATE Let $\tau=t=0$, $\varphi_{k}^{\left(0\right)}=\phi_{k}^{\left(0\right)}=\bar{\gamma}_{k}$ and $\psi_{k}^{\left(0\right)}=\widetilde{V}\left(\phi_{k}^{\left(0\right)}\right)$, $\forall k\in\mathcal{K}$. Calculate the objective value $\xi^{\left(\tau\right)}$ and $\varsigma^{\left(t\right)}$.\label{URLLCA0103}
\STATE Let $t\leftarrow t+1$. Solve problem~\eqref{URLLC41} to obtain $\xi^{\left(t\right)}$, $q_{k}^{\left(t\right)}$, $\varphi_{k}^{\left(t\right)}$, $\psi_{k}^{\left(t\right)}$, $\phi_{k}^{\left(t\right)}$, $\theta_{k}^{\left(t\right)}$, $a_{k,l}^{\left(t\right)}$, and $b_{k,l}^{\left(t\right)}$,
$\forall k,l\in\mathcal{K}$, $k\neq l$, with $\mathbf{w}_{k}^{\left(l\right)}$.\label{URLLCA0104}
\STATE If $\left|\frac{\varsigma^{\left(t\right)}-\varsigma^{\left(t-1\right)}}{\varsigma^{\left(t-1\right)}}\right|\leq\varepsilon$, stop iteration. Otherwise, go to Step~\ref{URLLCA0104}.  \label{URLLCA0105}
\STATE Let $\tau=\tau+1$, update $\mathbf{w}_{k}^{\left(\tau\right)}$ with $q_{k}^{\left(t\right)}$ and~\eqref{URLLC33}, and calculate the objective value $\xi^{\left(\tau\right)}$. If $\left|\frac{\xi^{\left(\tau\right)}-\xi^{\left(\tau-1\right)}}{\xi^{\left(\tau-1\right)}}\right|\leq\varepsilon$, stop iteration. Otherwise, go to Step~\ref{URLLCA0104}.\label{URLLCA0106}
\end{algorithmic}
\end{algorithm}

\subsection{For EE Maximization}

Different from the SRMax problem~\eqref{URLLC05}, the objective function of EEMax problem~\eqref{URLLC06} is in the fractional form, which makes the problem more challenging to obtain the global optimal solution. It is not difficult to find that the optimal beamforming vector $\left\{\mathbf{w}_{k}\right\}$ is calculated by~\eqref{URLLC33} when the transmitting power $\left\{q_{k}\right\}$ is fixed. Similar to problem~\eqref{URLLC34}, given the beamforming vector $\left\{\mathbf{w}_{k}\right\}$, problem~\eqref{URLLC31} is reformulated as
\begin{subequations}\label{URLLC44}
\begin{align}
&\max_{\left\{q_{k}\right\}}~\frac{\sum\limits_{k\in\mathcal{K}}\alpha_{k}\overleftarrow{R}\left(\overleftarrow{\gamma}_{k}\right)}
{\eta\sum\limits_{k\in\mathcal{K}}q_{k}+N_{\mathrm{t}}P_{\mathrm{c}}+P_{0}},\label{URLLC44a}\\
\mathrm{s.t.}~&\sum\limits_{k\in\mathcal{K}}q_{k}\leq P,~\nu_{3}\leq\overleftarrow{\gamma}_{k},~\forall k\in\mathcal{K}. \label{URLLC44b}%
\end{align}
\end{subequations}
To efficiently address problem~\eqref{URLLC44}, similarly, we rewrite it as the following equivalent problem:
\begin{subequations}\label{URLLC45}
\begin{align}
&\max~\frac{\sum\limits_{k\in\mathcal{K}}\alpha_{k}\tilde{R}\left(\varphi_{k},\theta_{k}\right)}
{\eta\sum\limits_{k\in\mathcal{K}}q_{k}+N_{\mathrm{t}}P_{\mathrm{c}}+P_{0}},\label{URLLC45a}\\
\mathrm{s.t.}~&\eqref{URLLC35b}-\eqref{URLLC35i}.\label{URLLC45b}
\end{align}
\end{subequations}
In problem~\eqref{URLLC45}, the optimization variables are $q_{k}$, $\varphi_{k}$, $\psi_{k}$, $\phi_{k}$, $\theta_{k}$, $a_{k,l}$, and $b_{k,l}$, $\forall k, l\in\mathcal{K}$, $k\neq l$.  Note that the objective function given in~\eqref{URLLC45a} is in concave-convex form. Exploiting the conclusion obtained in~\cite{BullDink1967}, we introduce parameter $\lambda$ to transform the objective function given in~\eqref{URLLC45a} into a parameterized subtractive form and use similar convex approximation to obtain
\begin{subequations}\label{URLLC46}
\begin{align}
\max~\sum\limits_{k\in\mathcal{K}}\alpha_{k}\tilde{R}\left(\varphi_{k},\theta_{k}\right)-\lambda
\left[\eta\sum\limits_{k\in\mathcal{K}}q_{k}+N_{\mathrm{t}}P_{\mathrm{c}}+P_{0}\right],
\mathrm{s.t.}~\eqref{URLLC41b}-\eqref{URLLC41j}.\label{URLLC46b}
\end{align}
\end{subequations}
In problem~\eqref{URLLC46}, the optimization variables are $q_{k}$, $\varphi_{k}$, $\psi_{k}$, $\phi_{k}$, $\theta_{k}$, $a_{k,l}$, and $b_{k,l}$, $\forall k, l\in\mathcal{K}$, $k\neq l$.  Now, exploiting the Dinkelbach method~\cite{BullDink1967}, an alternative optimization algorithm can be designed to address problem~\eqref{URLLC45}. The detailed steps of solving problem~\eqref{URLLC31} is outlined in Algorithm~\ref{URLLCA02} where $\tilde{\xi}^{\left(\tau\right)}$ and $\tilde{\varsigma}^{\left(t\right)}$ denote the objective values of problem~\eqref{URLLC31} and problem~\eqref{URLLC46} at the $\tau$-th and $t$-th iterations, respectively.
\begin{algorithm}[htp]
\caption{Solution of problem~\eqref{URLLC31}}\label{URLLCA02}
\begin{algorithmic}[1]
\STATE Let $\lambda=0$;\label{URLLCA0201}
\STATE Initialize beamforming vector $\mathbf{w}_{k}^{\left(0\right)}$ and $p_{k}^{\left(0\right)}$, $\forall k\in\mathcal{K}$, such that constraint~\eqref{URLLC06b} and~\eqref{URLLC06c} are satisfied. \label{URLLCA0202}
\STATE Compute $\gamma_{k}$ with $\mathbf{w}_{k}^{\left(0\right)}$ and $p_{k}^{\left(0\right)}$ to obtain $\bar{\gamma}_{k}$. Compute $\tilde{q}_{k}$ that is the $k$-th element of vector $\tilde{\mathbf{q}}=\left[\tilde{q}_{1},\cdots,\tilde{q}_{K}\right]^{T}$ with~\eqref{URLLC42} and~\eqref{URLLC43}. \label{URLLCA0203}
\STATE Let $\tau=t=0$, $\varphi_{k}^{\left(0\right)}=\phi_{k}^{\left(0\right)}=\bar{\gamma}_{k}$ and $\psi_{k}^{\left(0\right)}=\widetilde{V}\left(\phi_{k}^{\left(0\right)}\right)$, $\forall k\in\mathcal{K}$. Calculate the objective value $\tilde{\xi}^{\left(t\right)}$ and $\tilde{\varsigma}^{\left(\tau\right)}$.\label{URLLCA0204}
\STATE Let $t\leftarrow t+1$. Solve problem~\eqref{URLLC46} to obtain $\tilde{\xi}^{\left(t\right)}$, $q_{k}^{\left(t\right)}$, $\varphi_{k}^{\left(t\right)}$, $\psi_{k}^{\left(t\right)}$, $\phi_{k}^{\left(t\right)}$, $\theta_{k}^{\left(t\right)}$, $a_{k,l}^{\left(t\right)}$, and $b_{k,l}^{\left(t\right)}$,
$\forall k,l\in\mathcal{K}$, $k\neq l$, with $\mathbf{w}_{k}^{\left(l\right)}$.\label{URLLCA0205}
\STATE If $\left|\frac{\tilde{\varsigma}^{\left(t\right)}-\tilde{\varsigma}^{\left(t-1\right)}}{\tilde{\varsigma}^{\left(t-1\right)}}\right|\leq\varepsilon$, stop iteration. Otherwise, go to Step~\ref{URLLCA0205}.  \label{URLLCA0206}
\STATE If $\left|\sum\limits_{k\in\mathcal{K}}\alpha_{k}\tilde{R}\left(\varphi_{k}^{\left(t\right)},\theta_{k}^{\left(t\right)}\right)-\lambda
\left[\eta\sum\limits_{k\in\mathcal{K}}q_{k}^{\left(t\right)}+N_{\mathrm{t}}P_{\mathrm{c}}+P_{0}\right]\right|<\varepsilon$, then stop iteration. Otherwise, let $\lambda=\frac{\sum\limits_{k\in\mathcal{K}}\alpha_{k}\tilde{R}\left(\varphi_{k}^{\left(t\right)},\theta_{k}^{\left(t\right)}\right)}
{\eta\sum\limits_{k\in\mathcal{K}}q_{k}^{\left(t\right)}+N_{\mathrm{t}}P_{\mathrm{c}}+P_{0}}$, go to Step~\ref{URLLCA0205}.\label{URLLCA0207}
\STATE Let $\tau=\tau+1$, update $\mathbf{w}_{k}^{\left(\tau\right)}$ with $q_{k}^{\left(t\right)}$ and~\eqref{URLLC33}, and calculate the objective value $\tilde{\xi}^{\left(\tau\right)}$. If $\left|\frac{\tilde{\xi}^{\left(\tau\right)}-\tilde{\xi}^{\left(\tau-1\right)}}{\tilde{\xi}^{\left(\tau-1\right)}}\right|\leq\varepsilon$, stop iteration. Otherwise, go to Step~\ref{URLLCA0205}.\label{URLLCA0208}
\end{algorithmic}
\end{algorithm}

\subsection{For Fairness Optimzation}
Similar to problems~\eqref{URLLC34} and~\eqref{URLLC44}, when the beamforming vector $\left\{\mathbf{w}_{k}\right\}$ is fixed, the MaxMin problem~\eqref{URLLC32} can be reformulated as
\begin{subequations}\label{URLLC47}
\begin{align}
&\max_{\left\{q_{k}\right\}}\min\limits_{k\in\mathcal{K}}~\overleftarrow{\gamma}_{k},\label{URLLC47a}\\
\mathrm{s.t.}~&\sum\limits_{k\in\mathcal{K}}q_{k}\leq P, \label{URLLC47b}\\
&\nu_{3}\leq \overleftarrow{\gamma}_{k}, \forall k\in\mathcal{K}. \label{URLLC47c}
\end{align}
\end{subequations}
Further, problem~\eqref{URLLC47} can be equivalently rewritten as follows
\begin{subequations}\label{URLLC48}
\begin{align}
&\max_{\left\{q_{k},\mu\right\}}~\mu,\label{URLLC48a}\\
\mathrm{s.t.}~&\sum\limits_{k\in\mathcal{K}}q_{k}\leq P, \label{URLLC48b}\\
&\nu_{3}\leq \overleftarrow{\gamma}_{k}, \forall k\in\mathcal{K}, \label{URLLC48c}\\
&\mu\leq\overleftarrow{\gamma}_{k}, \forall k\in\mathcal{K}. \label{URLLC48d}
\end{align}
\end{subequations}
It is not difficult to find that problem~\eqref{URLLC48} can be reformulated as a standard geometric programming (GP) problem by jointly using the logarithmic change of the
variable and a logarithmic transformation of the objective function and constraints, respectively~\cite{GPBoyd2007}. In other words, problem~\eqref{URLLC48} can be easily solved by using the powerful GP optimization tool packets~\cite{Mosek2011}. The detailed algorithm for solving problem~\eqref{URLLC32} is summarized as Algorithm~\ref{URLLCA03}, where $\bar{\xi}^{\left(t\right)}$ denotes the objective value of problem~\eqref{URLLC32} at the $t$-th iteration. The convergence of Algorithm~\ref{URLLCA03} can be guaranteed using the monotonic boundary theorem~\cite{Bibby1974}. This is because that a monotonic nondecreasing sequence is generated with the running of Algorithm~\ref{URLLCA03}.
\begin{algorithm}[htp]
\caption{Solution of problem~\eqref{URLLC32}}\label{URLLCA03}
\begin{algorithmic}[1]
\STATE Initialize beamforming vector $\mathbf{w}_{k}^{\left(0\right)}$ and $p_{k}^{\left(0\right)}$, $\forall k\in\mathcal{K}$, such that constraints~\eqref{URLLC47b} and~\eqref{URLLC47c} are satisfied. \label{URLLCA0301}
\STATE Compute $\gamma_{k}$ with $\mathbf{w}_{k}^{\left(0\right)}$ and $p_{k}^{\left(0\right)}$ to obtain $\bar{\gamma}_{k}$. Compute $\tilde{q}_{k}$ that is the $k$-th element of vector $\tilde{\mathbf{q}}=\left[\tilde{q}_{1},\cdots,\tilde{q}_{K}\right]^{T}$ with~\eqref{URLLC42} and~\eqref{URLLC43}. \label{URLLCA0302}
\STATE Let $t=0$. Calculate the objective value $\bar{\xi}^{\left(t\right)}$.\label{URLLCA0303}
\STATE Let $t\leftarrow t+1$. Solve problem~\eqref{URLLC48} to obtain $\mu^{\left(t\right)}$, $q_{k}^{\left(t\right)}$, $\forall k\in\mathcal{K}$, with $\mathbf{w}_{k}^{\left(t\right)}$.\label{URLLCA0304}
\STATE Update $\mathbf{w}_{k}^{\left(t\right)}$ with $q_{k}^{\left(t\right)}$ and~\eqref{URLLC33}, and calculate the objective value $\bar{\xi}^{\left(t\right)}$. If $\left|\frac{\bar{\xi}^{\left(t\right)}-\bar{\xi}^{\left(t-1\right)}}{\bar{\xi}^{\left(t-1\right)}}\right|\leq\varepsilon$, stop iteration. Otherwise, go to Step~\ref{URLLCA0304}.\label{URLLCA0305}
\end{algorithmic}
\end{algorithm}

\subsection{Feasibility Analysis and Initialization}
To effectively solve problems~\eqref{URLLC05} and~\eqref{URLLC06}, another thing that needs to be addressed is to investigate their feasibility. According to the conclusion obtained in Theorem~\ref{URLLCTheorem02}, the value of SINR $\gamma_{k}$ must be greater than $\nu_{3}$, $\forall k\in\mathcal{K}$. In what follows, we resort to address the power minimization optimization formulated in~\eqref{URLLC49} to discuss their feasibility.
\begin{subequations}\label{URLLC49}
\begin{align}
&\min\limits_{\left\{\tilde{\mathbf{w}}_{k}\right\}}~\sum\limits_{k\in\mathcal{K}}\left\|\tilde{\mathbf{w}}_{k}\right\|^{2}\\
\mathrm{s.t.}~&\nu_{3}\leq\frac{\left|\overline{\mathbf{h}}_{k}^{H}\tilde{\mathbf{w}}_{k}\right|^{2}}
{\sum\limits_{l\neq k}\left|\overline{\mathbf{h}}_{k}^{H}\tilde{\mathbf{w}}_{l}\right|^{2}+1}, \forall k\in\mathcal{K}.
\end{align}
\end{subequations}
As the phases of the beamforming vectors $\left\{\tilde{\mathbf{w}}_{k}\right\}$ do not change the objective nor the constraints. Therefore, without loss of generality, we assume that the beamforming vectors $\left\{\tilde{\mathbf{w}}_{k}\right\}$ satisfy $\overline{\mathbf{h}}_{k}^{H}\tilde{\mathbf{w}}_{k}> 0$~\cite{TSPWiesel2006}. Let $\mathbf{o}=\left[\overline{\mathbf{h}}_{k}^{H}\tilde{\mathbf{w}}_{1},\overline{\mathbf{h}}_{k}^{H}\tilde{\mathbf{w}}_{2},\cdots,
\overline{\mathbf{h}}_{k}^{H}\tilde{\mathbf{w}}_{K-1},\overline{\mathbf{h}}_{k}^{H}\tilde{\mathbf{w}}_{K},1\right]$. Thus, problem~\eqref{URLLC49} can be rewritten as follows:
\begin{subequations}\label{URLLC50}
\begin{align}
&\min\limits_{\left\{\tilde{\mathbf{w}}_{k}\right\}}~\sum\limits_{k\in\mathcal{K}}\left\|\tilde{\mathbf{w}}_{k}\right\|^{2},~
\mathrm{s.t.}~\left\|\mathbf{o}\right\|\leq\sqrt{1+\frac{1}{\nu_{3}}}\overline{\mathbf{h}}_{k}^{H}\tilde{\mathbf{w}}_{k}.
\end{align}
\end{subequations}
Problem~\eqref{URLLC50} can be easily solved by the classical convex optimization tools, such as CVX~\cite{CVXTool}. Let $\left\{\tilde{\mathbf{w}}_{k}^{\left(*\right)}\right\}$ and $p^{\left(*\right)}$ be the optimum solution and target value of problem~\eqref{URLLC49}, respectively. If $p^{\left(*\right)}\leq P$, problems~\eqref{URLLC05},~\eqref{URLLC06}, and~\eqref{URLLC07} are feasible. Otherwise, the two problems are not feasible. Furthermore, the initialization of $\left\{p_{k}\right\}$ and $\left\{\mathbf{w}_{k}\right\}$ can be achieved by letting $p_{k}^{\left(0\right)}=\left\|\tilde{\mathbf{w}}_{k}^{\left(*\right)}\right\|^{2}$ and $\mathbf{w}_{k}^{\left(0\right)}=\frac{\tilde{\mathbf{w}}_{k}^{\left(*\right)}}{\left\|\tilde{\mathbf{w}}_{k}^{\left(*\right)}\right\|}$, $\forall k\in\mathcal{K}$.

\section{\label{simulation} Numerical Results}
In this section, we present numerical results to evaluate the performance of the proposed algorithms for the downlink  multiuser uRLLC communication system with finite blocklength transmission. Our simulation model consists of a single multi-antenna BS and $K$ single antenna users.  The channel coefficient $\mathbf{h}_{k}$ from the BS to the $k$-th user is modeled as $\mathbf{h}_{k}=\sqrt{\varrho_{k}}\widetilde{\mathbf{h}}_{k}$, where the channel power $\varrho_{k}$ is given as $\varrho_{k} = 1/\left(1 + (d_{k}/d_{0}\right)^{\varrho})$ with $d_{k}$ being the distance between the BS and the $k$-th user, $d_{0}$ and $\varrho$ respectively denoting the reference distance and the fading exponent. The elements of $\widetilde{\mathbf{h}}_{k}$ are independent and identically distributed (i.i.d.) with $\mathcal{CN}\left(0,1\right)$. All users have the same noise variance, i.e., $\sigma_{k}^{2}=\sigma^{2}$, $\forall k\in\mathcal{K}$.  For easy of notation, we define the SNR as $\mathrm{SNR}=10\log_{10}\left(\frac{P}{\sigma^{2}}\right)$ in dB. Other simulation parameters are listed in detail in Table~\ref{SimulationPara}. 
\begin{table*}[htbp]
\renewcommand{\captionfont}{\footnotesize}
\setlength{\abovecaptionskip}{0pt}
\setlength{\belowcaptionskip}{5pt}
\centering
\caption{Simulation parameters.}
\begin{tabular}{|c|c|}
\hline
$\varrho$ & $3$  \\ \hline
Reference distance $d_{0}$ & $50$ m  \\ \hline
Cell radius & $300$ m \\ \hline
Users distribution & Randomly distributed between $d_0$ and cell boundary  \\ \hline
Small scale fading distribution & Rayleigh fading with unit variance  \\ \hline
Circuit power per antenna $P_{\mathrm{c}}$ & $30$ dBm  \cite{TSPHe2014}\\ \hline
Static circuit power consumption $P_{\mathrm{0}}$ & $40$ dBm \cite{TSPHe2014}  \\ \hline
Number of bits to be sent $D$ & $32$ bytes \\ \hline
Bit error rate $\epsilon$ & $10^{-2}-10^{-10}$ \\ \hline
BS antenna number $N_{\mathrm{t}}$& $32$\\ \hline
Range of SNR & $15-30$ dB\\ \hline
Range of user number $K$ & $4-16$\\ \hline
Blocklength $n$ & $32$, $64$, $128$, $256$, $512$\\ \hline
\end{tabular}
\label{SimulationPara}
\end{table*}

To the best of our knowledge, there exists no published work on joint power allocation and coordinated beamforming for multi-antenna communication systems with finite blocklength transmission. We could only compare the proposed SRMax, EEMax and MaxMin with the conventional work in infinite blocklength. For example, the weighted SRMax problem is formulated as follows
\begin{subequations}\label{URLLC51}
\begin{align}
&\max_{\left\{\mathbf{w}_{k}, p_{k}\right\}}~\sum\limits_{k\in\mathcal{K}}\alpha_{k}C\left(\gamma_{k}\right),\label{URLLC51a}\\
\mathrm{s.t.}~&\sum\limits_{k\in\mathcal{K}}p_{k}\leq P, \left\|\mathbf{w}_{k}\right\|_{2}^{2}=1, \forall k\in\mathcal{K}, \label{URLLC51b}\\
&\frac{D}{n}\ln\left(2\right)\leq C\left(\gamma_{k}\right), \forall k\in\mathcal{K}. \label{URLLC51c}
\end{align}
\end{subequations}
Similarly, the MaxMin problem with infinite  blocklength can be formulated as
\begin{subequations}\label{URLLC52}
\begin{align}
&\max_{\left\{\mathbf{w}_{k}, q_{k}\right\}}\min\limits_{k\in\mathcal{K}}~{\gamma}_{k},\label{URLLC52a}\\
\mathrm{s.t.}~&\sum\limits_{k\in\mathcal{K}}q_{k}\leq P, \left\|\mathbf{w}_{k}\right\|_{2}^{2}=1, \forall k\in\mathcal{K},  \label{URLLC52b}\\
&\frac{D}{n}\ln\left(2\right)\leq C\left({\gamma_{k}}\right), \forall k\in\mathcal{K}. \label{URLLC52c}
\end{align}
\end{subequations}
Note that the problems~\eqref{URLLC51} and~\eqref{URLLC52} can be easily solved with similar methods as described in Algorithm~\ref{URLLCA01} and Algorithm~\ref{URLLC03}. In addition, the initializations of problems~\eqref{URLLC51} and~\eqref{URLLC52} can be realized via solving~\eqref{URLLC49} with replacing $\nu_{3}$ with $\tilde{\nu}_{3}=2^{\frac{D}{n}}-1$.

\subsection{\label{feasibility_simulation} Feasibility and initialization experiment}

Experiment~\ref{feasibility_simulation} discusses the performance of the proposed initialization algorithm as described in~\eqref{URLLC49}. In the experiment, channel coefficients are first generated with different transmitting parameters, then the power minimization algorithm~\eqref{URLLC49} is applied to judge their feasibilities. Monte Carlo method is adopted to obtain the feasible probability of the proposed power minimization algorithm~\eqref{URLLC49} (denoted as $Pr_{(\ref{URLLC49})}$). The feasible probabilities of ZFBF and solving the power minimization algorithm~\eqref{URLLC49} with replacing $\nu_{3}$ with $\tilde{\nu}_{3}=2^{\frac{D}{n}}-1$ for initializing problem~\eqref{URLLC51} are also listed, denoted as $Pr_{\rm{ZFBF}}$ and $Pr_{(\ref{URLLC51})}$, respectively. The feasible probability is defined as
\begin{subequations}\label{URLLC53}
\begin{align}
&Pr  = \frac{{\sum\limits_{i = 1}^M {f_{i}({\gamma _k})} }}{M},\nonumber\\
&f_{i}(\gamma _{k}) = \prod\limits_{k \in \mathcal{K}} {I({\gamma _k} - {\upsilon _3})},\nonumber
\end{align}
\end{subequations}
where $M$ is the Monte Carlo times, $\gamma_k$ is the  calculated SINR of the $k$-th user. $I(x)$ is an index function, i.e., $I(x) = 1$ if $x\geq0$, otherwise $I(x) = 0$.

\begin{table}[htbp]
  \centering
  \caption{Feasible probability results of different algorithms}
    \begin{tabular}{|c|c|c|c|c|c|} 
    \hline
           \multicolumn{1}{|c|}{SNR(dB)} & \multicolumn{1}{c|}{$\epsilon$} & \multicolumn{1}{c|}{$K$} & \multicolumn{1}{c|}{$Pr_{(\ref{URLLC49})}$} & \multicolumn{1}{c|}{$Pr_{\rm{ZFBF}}$} & $Pr_{(\ref{URLLC51})}$ \\
          \hline
    \multirow{5}[3]{*}{20} & \multirow{5}[3]{*}{$10^{-5}$} & 2     & 100\% & 100\% & 100\% \\
    \cline{3-6}       &       & 4     & 98.50\% & 54.00\% & 49.40\% \\
    \cline{3-6}       &       & 6     & 84.00\% & 12.40\% & 7.50\% \\
    \cline{3-6}       &       & 8     & 55.70\% & 1.70\% & 1.20\% \\
    \cline{3-6}       &       & 10    & 22.20\% & 0.50\% & 0.30\% \\
    \cline{1-6} 25    & \multirow{3}[2]{*}{$10^{-5}$} & \multirow{3}[2]{*}{6} & 100\% & 100\% & 100\% \\
    \cline{1-1}\cline{4-6}  20    &       &       & 84.00\% & 12.40\% & 7.50\% \\
    \cline{1-1}\cline{4-6} 15    &       &       & 7.20\% & 0.60\% & 0.50\% \\
    \cline{1-1}\cline{2-6} \multirow{3}[3]{*}{20} & $10^{-2}$ & \multirow{3}[2]{*}{6} & 98.00\% & 20.40\% & 17.50\% \\
    \cline{2-2} \cline{4-6}      & $10^{-4}$ &       & 87.50\% & 14.70\% & 13.50\% \\
    \cline{2-2} \cline{4-6}      & $10^{-6}$ &       & 79.00\% & 11.10\% & 9.10\% \\
    \hline
    \end{tabular}%
  \label{tab:Feasibility}%
\end{table}%

Table \ref{tab:Feasibility} lists the feasible probabilities of three methods, where $N_{\mathrm{t}}$ and $n$ are $32$ and $128$, respectively. Obviously, changes in three factors, including the increase of user number, the decrease of SNR and the increase of QoS requirement, will all lead to the decrease of feasible probability. Meanwhile, the values of $Pr_{\rm{ZFBF}}$ and $Pr_{\eqref{URLLC51}}$ are lower than that of $Pr_{\eqref{URLLC49}}$. Both conventional methods achieve only less than $15\%$ feasible probability, while it is
higher than $80\%$ of $Pr_{\eqref{URLLC49}}$, with the parameters $K = 6$, SNR = $20$ dB and $\epsilon = 10^{-5}$. It is worth noting that, $Pr_{\eqref{URLLC49}}$ is still less than $100\%$ with parameters $K = 4$, $\epsilon = 10^{-5}$ and SNR $= 20$ dB. Infeasible instances may happen when the randomly generated users located at the edge of the cell. 

\subsection{\label{SRMax_experiment} SRMax optimization experiment}

This experiment discusses the performance of Algorithm~\ref{URLLCA01} with different number of users and blocklengths with $\alpha_{k} = \frac{1}{K}$, $N_\mathrm{t}= 32$, SNR = $20$ dB, and $\epsilon = 10^{-5}$. Channel coefficients with parameters $K = 4, 6, 8, 10$, $n = 128$ are first generated and then Algorithm~\ref{URLLCA01} is implemented. Results of ZFBF will also be illustrated for comparison. The weighted sum rate for ZFBF is the same as illustrated in~\eqref{URLLC05a}. Before comparisons, Rectified Linear Unit (ReLU) below is used to process ZFBF results, i.e., $R({\gamma _k}) =\max\left(0,R\left(\gamma _{k}\right)\right)$, $k = 1,2,...,K$, since it may face infeasible problem as discussed in~\ref{feasibility_simulation}.
\begin{figure}[!htbp]
\renewcommand{\captionfont}{\footnotesize}
	\centering
	\includegraphics[width=0.5\columnwidth]{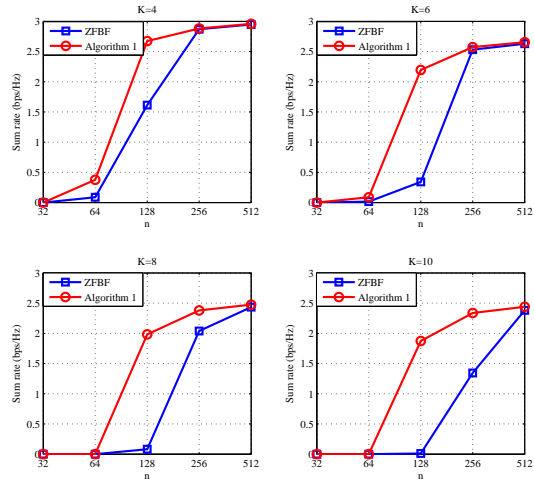}\\
	\caption{Sum rate with different blocklength and number of users.}
	\label{SR_param1}
\end{figure}

Fig.~\ref{SR_param1} shows the weighted sum rate results of Algorithm~\ref{URLLCA01} and ZFBF, respectively. It is clear that, if the blocklength is too small ($n \leq 32$), both algorithms could not obtain feasible results. With the increase of blocklength, both methods obtain a weighted sum rate, and they tend to be stable when $n\geq 256$. Meanwhile, with the increase of $K$, the weighted sum rate decreases since too many users compete for limited resources. On the other hand, Algorithm~\ref{URLLCA01} always outperforms ZFBF, especially when the blocklength is small or the number of users is large, which implies the effectiveness of Algorithm~\ref{URLLCA01} under extreme transmitting conditions.

\subsection{\label{EEMax} EEMax optimization experiment}
\begin{figure}[htbp]
\renewcommand{\captionfont}{\footnotesize}
	\centering
	\includegraphics[width=0.5\columnwidth,keepaspectratio]{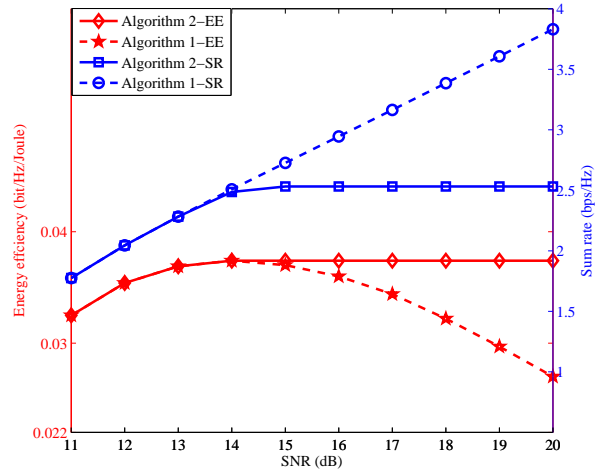}\\
	\caption{Sum rate and energy efficiency with different SNRs.}
	\label{EE_param2}
\end{figure}

In this section, the energy efficiency performance using Algorithm~\ref{URLLCA02} through Monte Carlo simulations is presented with $\epsilon = 10^{-5}$, $K = 6$, and $n = 128$. Channel coefficients under the worst SNR condition (SNR = $11$ dB) are first generated to confirm the feasibility. Then Algorithm~\ref{URLLCA02} is implemented and energy efficiency curve (Algorithm~\ref{URLLCA02}-EE) could be first plotted with different SNRs in Fig.~\ref{EE_param2}. In order to investigate the relationship between energy efficiency and sum rate, the sum rate result is also recorded while energy efficiency is optimized. Sum rate curve (Algorithm~\ref{URLLCA02}-SR) is synchronously plotted in the same figure. On the other hand, the sum rate results (Algorithm~\ref{URLLCA01}-SR) and energy efficiency (Algorithm~\ref{URLLCA01}-EE) results using Algorithm~\ref{URLLCA01} on the same data are also plotted in Fig.~\ref{EE_param2} to explore the difference between the two algorithms.

In the figure, the energy efficiency of Algorithm~\ref{URLLCA02}-EE increases while $11$ dB $<=$ SNR$<=$ $14$ dB. It reaches the maximum value SNR$ = 14$ dB and then tends to be unchanged when SNR increases to $20$ dB, which means that Algorithm~\ref{URLLCA02} will not consume more power if the maximum energy efficiency is obtained. Therefore, Algorithm~\ref{URLLCA02}-SR has the similar trends with Algorithm~\ref{URLLCA02}-EE. On the other hand, Algorithm~\ref{URLLCA01} aims to maximize the weighted sum rate at the expense of energy efficiency. Therefore, Algorithm~\ref{URLLCA01}-EE increases briefly while SNR increases from $11$ dB to $14$ dB, and then decreases while SNR increases to $20$ dB.

\begin{figure}[!htbp]
\renewcommand{\captionfont}{\footnotesize}
	\centering
	\includegraphics[width=0.5\columnwidth,keepaspectratio]{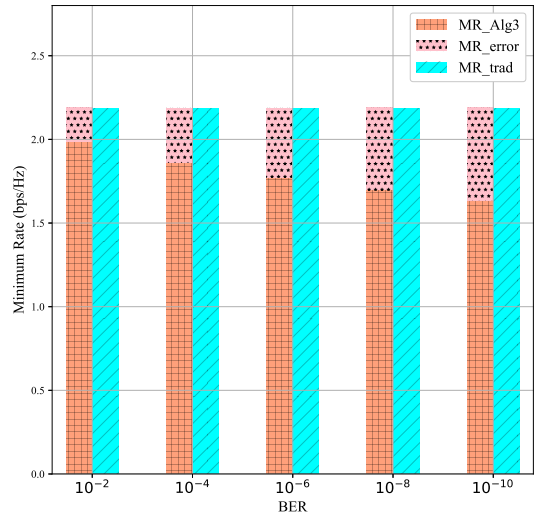}\\
	\caption{Minimum rate with different BERs.}
	\label{SR_param3}
\end{figure}
\subsection{\label{MaxMin} Fairness optimization experiment}

Experiment~\ref{MaxMin} discusses performance of Algorithm~\ref{URLLCA03} under different QoS requirements. Traditional processing method using Shannon capacity, as the problem studied in~\eqref{URLLC52} is also compared with $K = 6$, $n = 128$, and SNR = 20 dB. Again, channel coefficients under the worst condition ($\epsilon = 10^{-10}$ ) are first generated before
the above two algorithms are implemented. The minimum rate for Algorithm~\ref{URLLCA03} (MR$\_$Alg3) and traditional method using Shannon capacity (MR$\_$trad), and an error term MR$\_$error are  respectively defined as follows:
\begin{subequations}
\begin{align}
&{\rm{MR}}\_{\rm{Alg}}3 = \log (1 + \gamma ) - \frac{{{Q^{ - 1}}(\epsilon )}}{{\sqrt n }}\sqrt {{{(1 - \frac{1}{{(1 + \gamma )}})}^2}}\nonumber \\
&{\rm{MR}}\_{\rm{trad = }}\log (1 + \gamma ) \nonumber\\
&{\rm{MR\_error = }}\frac{{{Q^{ - 1}}(\varepsilon )}}{{\sqrt n }}\sqrt {{{(1 - \frac{1}{{(1 + \gamma )}})}^2}}\nonumber
\end{align}
\end{subequations}

In the experiment, MR$\_$Alg3 and MR$\_$trad are directly calculated using the obtained beamforming via running Algorithm~\ref{URLLCA03} and solving problem~\eqref{URLLC52}, respectively. MR$\_$error is calculated using $\gamma$ which is obtained by using the obtained beamforming via running Algorithm~\ref{URLLCA03}. Fig.~\ref{SR_param3} illustrates results of  MR$\_$Alg3, MR$\_$trad and MR$\_$error. MR$\_$Alg3 decreases with an decreasing $\epsilon$, which indicates that more strict QoS requirement reduces transmission capacity. On the contrary, MR$\_$trad remains unchanged since conventional method ignores the error term which could not be eliminated under finite blocklength circumstance. Therefore, MR$\_$trad is not an accessible result in practice compared to MR$\_$Alg3. Relationship among MR$\_$Alg3, MR$\_$error and MR$\_$trad, namely, summation of MR$\_$Alg3 and MR$\_$error is close to MR$\_$trad in theory, can also be experimentally confirmed in Fig.~\ref{SR_param3}.

\section{\label{conclusion} Conclusions}

In this paper, for the downlink uRLLC multiuser multi-antenna system, we focused on the beamforming design respectively for the weighted sum rate maximization, energy efficiency maximization, and fairness optimization by considering the maximum allowable transmission power and minimum rate requirement. We also analyzed the feasibility of the formulated optimization problem and the feasibility condition avoiding the situation where the rate of one of the users is zero. Using some basic mathematical operations, successive convex approximation method, and the uplink-downlink duality theory, we provided effective and efficient optimization algorithms to solve the formulated problems. Simulation results show that our proposed results outperforms ZFBF with equal power allocation, especially for the case with small transmission power.

\begin{small}

\end{small}
\end{document}